\documentclass[longauth]{aa}
\usepackage{txfonts}
\usepackage{graphicx}
\usepackage{natbib}
\bibpunct{(}{)}{;}{a}{}{,} 

\begin{document}

\newcommand{\Cerenkov}{Cherenkov\ }
\newcommand{\HMS}[3]{$#1^{\mathrm{h}}#2^{\mathrm{m}}#3^{\mathrm{s}}$}
\newcommand{\DMS}[3]{$#1^\circ #2' #3"$}
\newcommand{\TODO}[1]{\textbf{\texttt{TODO: \emph{#1}}}}
\newcommand{\UPDATED}[1]{#1}
\newcommand{\REMOVED}[1]{}
\renewcommand{\cite}{\citep} 

\newcommand{\StatSysErr}[3]{$#1 \pm #2_{\mathrm{stat}} \pm #3_{\mathrm{sys}}$}

\newcommand{\HESSa}{HESS~J1427$-$608}
\newcommand{\HESSb}{HESS~J1626$-$490}
\newcommand{\HESSc}{HESS~J1702$-$420}
\newcommand{\HESSd}{HESS~J1708$-$410}
\newcommand{\HESSe}{HESS~J1731$-$347}
\newcommand{\HESSf}{HESS~J1841$-$055}
\newcommand{\HESSg}{HESS~J1857$+$026}
\newcommand{\HESSh}{HESS~J1858$+$020}

\title{HESS VHE Gamma-Ray Sources Without Identified Counterparts}

\titlerunning{Unidentified HESS Sources}
\authorrunning{The HESS Collaboration}

\author{F. Aharonian\inst{1,13}
 \and A.G.~Akhperjanian \inst{2}
 \and U.~Barres de Almeida \inst{8} \thanks{supported by CAPES Foundation, Ministry of Education of Brazil}
 \and A.R.~Bazer-Bachi \inst{3}
 \and B.~Behera \inst{14}
 \and M.~Beilicke \inst{4}
 \and W.~Benbow \inst{1}
 \and K.~Bernl\"ohr \inst{1,5}
 \and C.~Boisson \inst{6}
 \and O.~Bolz \inst{1}
% ...
}

\institute{
Max-Planck-Institut f\"ur Kernphysik, P.O. Box 103980, D 69029
Heidelberg, Germany
\and
 Yerevan Physics Institute, 2 Alikhanian Brothers St., 375036 Yerevan,
Armenia
\and
Centre d'Etude Spatiale des Rayonnements, CNRS/UPS, 9 av. du Colonel Roche, BP
4346, F-31029 Toulouse Cedex 4, France
\and
Universit\"at Hamburg, Institut f\"ur Experimentalphysik, Luruper Chaussee
149, D 22761 Hamburg, Germany
\and
Institut f\"ur Physik, Humboldt-Universit\"at zu Berlin, Newtonstr. 15,
D 12489 Berlin, Germany
\and
LUTH, Observatoire de Paris, CNRS, Universit\'e Paris Diderot, 5 Place Jules Janssen, 92190 Meudon, 
France
\and
DAPNIA/DSM/CEA, CE Saclay, F-91191
Gif-sur-Yvette, Cedex, France
\and
University of Durham, Department of Physics, South Road, Durham DH1 3LE,
U.K.
\and
Unit for Space Physics, North-West University, Potchefstroom 2520,
    South Africa
\and
Laboratoire Leprince-Ringuet, Ecole Polytechnique, CNRS/IN2P3,
 F-91128 Palaiseau, France
\and 
Laboratoire d'Annecy-le-Vieux de Physique des Particules, CNRS/IN2P3,
9 Chemin de Bellevue - BP 110 F-74941 Annecy-le-Vieux Cedex, France
\and
Astroparticule et Cosmologie (APC), CNRS, Universite Paris 7 Denis Diderot,
10, rue Alice Domon et Leonie Duquet, F-75205 Paris Cedex 13, France
\thanks{UMR 7164 (CNRS, Universit\'e Paris VII, CEA, Observatoire de Paris)}
\and
Dublin Institute for Advanced Studies, 5 Merrion Square, Dublin 2,
Ireland
\and
Landessternwarte, Universit\"at Heidelberg, K\"onigstuhl, D 69117 Heidelberg, Germany
% ...
}

\offprints{K. Kosack, \email{Karl.Kosack@mpi-hd.mpg.de}}

\abstract {The detection of gamma rays in the very-high-energy (VHE)
  energy range (100 GeV--100 TeV) provides a direct view of the parent
  population of ultra-relativistic particles found in astrophysical
  sources. For this reason, VHE gamma rays are useful for
  understanding the underlying astrophysical processes in non-thermal
  sources.}{We investigate unidentified VHE gamma-ray sources that
  have been discovered with HESS in the most sensitive blind survey of the
  Galactic plane at VHE energies conducted so far.}{The
  HESS array of imaging atmospheric Cherenkov telescopes (IACTs)
  has a high sensitivity compared with previous instruments
  ($\sim0.01\:\mathrm{Crab}$ in 25 hours observation time for a $5\sigma$
  point-source detection), and with its large field of view, is well suited for
  scan-based observations. The on-going HESS survey of the inner
  Galaxy has revealed a large number of new VHE sources, and for each
  we attempt to associate the VHE emission with multi-wavelength data
  in the radio through X-ray wavebands.  }{ For each of the eight
  unidentified VHE sources considered here, we present the energy
  spectra and sky maps of the sources and their environment. The VHE
  morphology is compared with available multi-wavelength data (mainly
  radio and X-rays). No plausible counterparts are found.}{}


\maketitle

\section{Introduction}

VHE gamma-ray astronomy has recently entered an new era of discovery
with the introduction of the latest generation \emph{Imaging
Atmospheric Cherenkov Telescopes} (IACTs) such as HESS (the High
Energy Stereoscopic System). Since HESS began operation in 2004,
about two dozen new VHE sources have been revealed. Presently
identified VHE gamma-ray sources belong to one of four categories:
active galactic nuclei (AGN), pulsar wind nebulae (PWN), shell-type
supernova remnants (SNR), or X-ray binaries (XRB); recently also VHE
emission was detected which may be associated with a young stellar
cluster \cite{HESS:Westerlund2}. All these identified source classes
also exhibit emission in the radio and/or X-ray regime. However,
several VHE sources discovered by HESS in the field-of-view of
other known sources \cite{HESS:J1303} or during the HESS Galactic
plane survey \cite{HESS:scanpaper1,HESS:scanpaper2} have not been
identified with objects from which VHE emission is expected. The first
unidentified VHE source was TeV~J2032+4130
\cite{HEGRA:TeVJ2032,HEGRA:TeVJ2032_Final}, which was discovered by
the HEGRA IACT system.  HESS~J1303-631 \cite{HESS:J1303} was found in
the field-of-view of the binary pulsar system PSR B1259-63/SS~2883,
and several other sources were subsequently discovered in the Galactic
plane survey. To date, these objects remain unidentified;
HESS~J1303-631 has even been postulated to be related to such an
exotic phenomenon as a gamma-ray burst remnant
\cite{atoyan06:HESSJ1303}.


VHE gamma rays are tracers of non-thermal particle acceleration, and
their production can be explained by the presence of either
high-energy electrons or protons. In electron scenarios, gamma rays
are primarily produced by inverse-Compton up-scattering of background
photon fields by high-energy electrons. Significant X-ray and radio
emission is predicted since the same population of electrons should
emit synchrotron radiation at \UPDATED{longer} wavelengths. For typical Galactic
magnetic field strengths, the energy flux of the X-ray component of
the photon spectrum in the keV range is predicted to be comparable to
the energy flux in the TeV range. The X-ray component of the
spectrum may be suppressed, however, if there is a cutoff in the
parent electron spectrum below $\sim$10 TeV
\cite{aharonian97:_gamma_x_ray_ratio}. In proton scenarios, VHE gamma
rays are produced primarily from the decay of neutral pions ($\pi^0$)
that result from proton-proton interactions. If gamma rays are
produced only via $\pi^0$ decay, a strong X-ray or radio signal may
not be present; however, proton interactions also produce charged
pions and cascades of secondary electrons that should generate a
continuum of X-ray and radio synchrotron emission. Since it is
difficult to explain VHE gamma-ray emission without at least a weak
lower-energy counterpart, the lack of low-energy emission from the
unidentified HESS sources puts significant constraints on physical
conditions and/or particle acceleration processes in their sources.
While the explanation may simply be that sufficiently deep
multi-wavelength observations of the objects have not yet been made,
the possibility exists that there is a new class of object that does
not follow the predictions of standard emission models.

Recent observations of the Galactic plane and further re-observations
of known sources with HESS have allowed for the study of some of
the weaker Galactic sources at increased sensitivity and have revealed
new VHE gamma-ray sources in addition to those described by
\citet{HESS:scanpaper2}. Similar to the previously mentioned objects,
several of these sources have no obvious cataloged counterpart at
\UPDATED{longer} wavelengths, and consequently their emission mechanism is
unidentified. In this paper, we focus on eight VHE emitters without
obvious counterpart that have been detected by HESS\UPDATED{.}  Of these
sources, an updated analysis is given for two previously published
unidentified sources for which subsequent observations have provided
significantly better statistics, and the detections of six new
unidentified sources are reported. New VHE detections within the
Galactic plane of known objects (PWN, SNRs, etc.) have been or will be
reported elsewhere
\cite[e.g. in][]{HESS:Monoceros,HESS:2PSR,HESS:Westerlund2}.

\section{Observations and Technique}

\subsection{The HESS Instrument}

HESS (the High Energy Stereoscopic System) is an array of four
atmospheric \Cerenkov telescopes located in the Khomas highland of
Namibia at an altitude of $1800\:\mathrm{m}$ above sea-level. Each
telescope consists of a $107 \mathrm{m^2}$ optical reflector made up
of segmented mirrors that focus light into a camera of 960
photo-multiplier tube pixels \cite{HESS:optics}. The telescopes image
the UV/blue flashes of \Cerenkov light emitted by the secondary
particles produced in gamma-ray-induced air-showers. Stereoscopic
shower observations using the \emph{imaging atmospheric \Cerenkov
technique} \citep[e.g.][]{hillas96:technique,weekes96:acts,HEGRA:acts}
allow for accurate reconstruction of the direction and energy of the
primary gamma rays as well as for the rejection of background events
from air showers of cosmic ray origin. HESS is sensitive to gamma
rays above a post-cuts threshold energy of approximately 150 GeV and
has an average energy resolution of $\sim16\%$
\cite{HESS:crab}. Additionally, the high angular resolution
($\sim{\hskip -0.3em}0.1^\circ$), large field-of-view ($\sim{\hskip
-0.3em}5^\circ$), and good off-axis sensitivity of the HESS array
make it well suited for extended sources and scan-based observations,
where the source position is not known a priori.


\subsection{Data}

The observations discussed here were taken as part of the ongoing
HESS Galactic plane survey which currently covers the band
$-50^\circ < l < 60^\circ$ in galactic longitude and $-3^\circ < b <
3^\circ$ in latitude.  Data were taken as a series of 28-minute
observations (runs) centered on regular grid points covering the survey
area. Additionally, several established sources were observed with
pointed follow-up observations in \emph{wobble mode}, where data are
taken with an alternating offset from the target position of typically
$\pm0.7^\circ$ in right ascension or declination.  The set of usable
runs were selected based on a standard set of hardware and weather
conditions \cite{HESS:crab}. The sources in this study were chosen by
selecting all locations in the HESS Galactic plane scan data set
that have a pre-trials detection significance (with a fixed
integration radius of 0.22$^\circ$) greater than 6$\sigma$
(corresponding to a post-trials significance of 4$\sigma$, based on
the very conservative estimate for the number of trials given in
\citet{HESS:scanpaper2}), and for which no obvious cataloged
counterpart can be associated (based on the criteria given in Section
\ref{sec:search}).  Sources that were previously published
\cite[e.g. in][]{HESS:scanpaper2} were excluded, except those that
have had increases in significance over 3 $\sigma$ due to subsequent
observation. The eight sources that pass these selection criteria and
their center positions (based on a model fit described in
\S\ref{sec:technique}) are summarized in Table \ref{tab:sources}. For
reference, a summary of published results on previously reported
unidentified VHE objects is given in Table \ref{tab:other_sources}.

\begin{table*}
  \begin{center}
    \begin{tabular}{l c c c c c c c}
      \hline\hline
      Source & Right Ascension & Declination & $l(^\circ)$ & $b(^\circ)$ & Time (hrs) & $S$ ($\sigma$) & Excess (cts)\\
      \hline
      \HESSa                   &  \HMS{14}{27}{52}  &  \DMS{-60}{51}{00}  & 314.409 & -0.145 & 21 & 7.3  & 197 \\
      \HESSb                   &  \HMS{16}{26}{04}  &  \DMS{-49}{05}{13}  & 334.772 & 0.045  & 12 & 7.5  & 153 \\
      \HESSc$\dagger$          &  \HMS{17}{02}{44}  &  \DMS{-42}{00}{57}  & 344.304 & -0.184 & 9 & 12.8 & 412 \\
      \HESSd$\dagger$          &  \HMS{17}{08}{24}  &  \DMS{-41}{05}{24}  & 345.683 & -0.469 & 39 & 10.7 & 513 \\
      \HESSe                   &  \HMS{17}{31}{55}  &  \DMS{-34}{42}{36}  & 353.565 & -0.622 & 14 & 8.1  & 218 \\
      \HESSf                   &  \HMS{18}{40}{55}  & \DMS{-05}{33}{00}    & 26.795  & -0.197 & 26 & 10.6 & 346 \\
      \HESSg                   &  \HMS{18}{57}{11}  &  \DMS{02}{40}{00}   & 35.972  & -0.056 & 21 & 8.7  & 223 \\
      \HESSh                   &  \HMS{18}{58}{20}  &  \DMS{02}{05}{24}   & 35.578  & -0.581 & 25 & 7.0  & 168 \\
      \hline\hline
      
    \end{tabular}
    \caption{ \label{tab:sources} Positions in equatorial (J2000
      epoch) and Galactic ($l$,$b$) coordinates along with the
      detection significances of unidentified sources in the
      HESS Galactic Plane scan discussed in this paper.  $S$ is
      the significance (number of standard deviations above the
      background level) of the source using a fixed integration radius
      of $0.22^\circ$, which was used for selecting the sources from
      the scan data. The position of each source is based on a model
      fit to the background-subtracted gamma-ray maps (discussed in
      \S\ref{sec:technique} and Table \ref{tab:morphology}). The fit
      positions have an average statistical error of 0.05
      degrees. Sources marked with a $\dagger$ are previously
      published in \citet{HESS:scanpaper2} and have been updated with
      new data. The exposure time is corrected for the off-axis
      sensitivity of the telescope system and accounts for instrumental readout
      dead-time. }
  \end{center}
\end{table*}

\begin{table}
  \begin{center}
    \begin{tabular}{l c c c}
      \hline\hline
      Source & R.A. & Dec & $\sigma_{\mathrm{src}}$ ($'$)\\
      \hline
      HESS~J1303-631$\ddag$ & \HMS{13}{03}{00}  & \DMS{-63}{11}{55}  & 9.6 \\
      HESS~J1614-518$\ddag$ & \HMS{16}{14}{19}  & \DMS{-51}{49}{12} &  13.8 \\
      HESS~J1632-478 & \HMS{16}{32}{09}  & \DMS{-47}{49}{12}  & 12.0\\ 
      HESS~J1634-472 & \HMS{16}{34}{58}  & \DMS{-47}{16}{12} & 6.6 \\ 
      HESS~J1745-303 & \HMS{17}{45}{02} & \DMS{30}{22}{12} & 12.6\\ 
      HESS~J1837-069 & \HMS{18}{37}{38} & \DMS{-6}{57}{00} & 7.2\\ 
      TeV~J2032+4130$\ddag$ & \HMS{20}{32}{57}  & \DMS{41}{29}{57} & 6.2 \\
      \hline\hline
    \end{tabular}    
    \caption{ \label{tab:other_sources} Previously published
      unidentified VHE sources, not discussed in this
      paper. Coordinates are in J2000 epoch, $\sigma_{\mathrm{src}}$
      is the intrinsic source extent (taking into account the
      instrumental response). Sources with $\ddag$ have no obvious
      \UPDATED{longer}-wavelength counterpart. HESS~J1632-478 has a possible HMXB
      counterpart, but the VHE source is extended; HESS~J1634-472 may
      be related to an unidentified INTEGRAL source or nearby SNR, but
      is offset and morphologically dissimilar; HESS~J1745-303 is
      partially coincident with an unidentified EGRET source; and
      HESS~J1837-069 is coincident with an as yet unidentified ASCA
      source. Results are from \citet{HESS:J1303},
      \citet{HESS:scanpaper2}, and \citet{HEGRA:TeVJ2032_Final}. }
  \end{center}
\end{table}

\subsection{Analysis Technique} 
\label{sec:technique}

The data presented here were analyzed using the standard
HESS analysis scheme: calibrations are applied to the raw shower
images \cite{HESS:calibration} followed by an image cleaning procedure
which removes noise due to fluctuations in the optical night-sky
background light. The images are then parametrized using the Hillas
moment-analysis technique \cite{hillas96:technique}, and gamma-ray
selection criteria based on the image parameters are applied
\cite{HESS:crab}. To reduce systematic effects in the spectrum due to
off-axis sensitivity that arise when images fall near the camera edge,
an additional cut is applied to accept only data runs which are taken
within an angular distance $\psi$ from the respective position of the
object under analysis. For the spectral analysis, this is
conservatively set to $2.0^\circ$ to minimize systematic errors on the
energy estimates (providing an average offset of $1.0^\circ \pm
0.1^\circ$), while for the generation of the sky maps, it was set to
$2.5^\circ$ to maximize the number of photons detected (giving an
average offset of $1.9^\circ \pm 0.2^\circ$).  Images from events
passing the cuts for each telescope are combined to reconstruct the
shower direction and energy. In the data presented here, two sets of
gamma-ray selection criteria are used to suppress events with hadronic
origin: \emph{standard cuts}, which are optimized using a simulated
source with an energy spectrum with photon index $\Gamma=2.6$ and a
flux that is 10\% of the Crab Nebula (a standard bright gamma-ray
source) at VHE energies, and \emph{hard cuts} which are optimized for
a harder spectrum source ($\Gamma=2.0$) with a flux that is 1\% of the
Crab Nebula. \emph{Standard cuts} have an intrinsically lower energy
threshold, but are looser and accept more background events, while the
\emph{hard cuts} provide better gamma-hadron separation, and thus
higher signal-to-noise ratio, at the expense of an increased energy
threshold. Unless otherwise noted, \emph{hard cuts} are employed for
the spectral and morphological analyses presented in this article
since they provide smaller systematic errors due to a higher analysis
energy threshold and better background rejection, though both sets are
applied to check for consistency.

The sky maps used for determining the source location and morphology
are generated by accumulating the points of origin of each gamma-ray
candidate in a two-dimensional histogram, subtracting a background map
modeled by counting the number of events which fall within an annulus
(of average radius $0.5^\circ$) about each grid point, excluding
emission regions \citep[the \emph{ring-background model} described
in][]{HESS:background}.  The background is corrected for acceptance
variations across the field of view. As an additional check, a
background model using the radial gamma-ray acceptance profile (as
determined by dedicated off-source observations and simulations) in
the field of view of each run is also used and compared for
consistency.  An elongated two-dimensional Gaussian convolved with the
HESS point-spread function is fit to the resulting excess map to
determine the centroid position, position angle, and extent of the
source. To define the full extent of the source for spectral analysis,
a histogram of the squared distance of each event to the fit position
($\theta^2$) is generated. The statistical
significances of each excess measurement are calculated from the
measured number of on- and off-source (background) events following
the likelihood ratio procedure outlined in \citet{li_ma83}.


The background for spectra is estimated using the
\emph{reflected-region technique} where background events are selected
from circular off-source regions in the field of view that have the
same angular size and offset from the observation center position as
the on-source region \cite{HESS:crab}. Background regions containing
other known sources are excluded. This technique provides a more
accurate estimation of the background than the field-of-view model
(described above) used to generate the sky maps, but is not as well
suited for the generation of two-dimensional images. 

Spectra are generated following \citet{HESS:crab} for all events that
fall within an angular distance $\theta_{\mathrm{int}}$ of the target
position. This radius is chosen for each source as the distance where
the radial excess distribution falls to a level indistinguishable from
noise (i.e. fully encloses the source). This provides a less biased
estimate of the spectrum since it makes no assumption on the source
morphology, but it decreases the signal-to-noise ratio since some
additional background is included compared to an angular cut optimized
for best significance. An energy estimate for each event is calculated
based on a comparison of the event's impact parameter, zenith angle,
offset from the center of the field of view, and the amplitude of the
integrated image for each telescope. The energy estimates for all
events in the on and off-source regions are put into two histograms,
which are then corrected for differing exposure, subtracted, and a
flux is calculated for each energy bin by dividing by the observation
time and the effective collection area of the telescopes (which is a
function of energy, offset from the camera center, zenith angle, and
the angle with respect to the Earth's geomagnetic field, as determined
from simulations). The resulting fluxes are fit by a power-law of the
form
\begin{equation}
  F(E) = N_0 \left(\frac{E}{1\:\mathrm{TeV}}\right)^{-\Gamma}
\end{equation}
where $\Gamma$ is the photon index and $N_0$ is the flux
normalization. Muon images are used to correct the energy estimate for
changes in the optical efficiency of the telescopes over time (due to,
e.g. the degradation of the mirrors) \cite{HESS:crab}. The systematic
error on the flux is conservatively estimated from simulated data to
be 20\% while the photon index has a typical systematic error of $\pm0.2$.


To check the robustness of the results presented in this article, the
analysis has been repeated using several other background models as
well as with a completely separate analysis and calibration chain
which used independent simulations and the \emph{forward-folded}
spectrum reconstruction technique described in
\citet{piron01:CAT_ForwardFolding_Mrk421}.


\subsection{Counterpart Search} \label{sec:search}

A search for counterparts to the VHE emission was made by first
looking in source catalogs for objects which are of a type known to
produce VHE photons, including the ATNF pulsar catalog \cite{ATNF},
the Green's supernova remnant catalog \cite{green04:SNRs}, and the
High-Mass X-ray binary (HMXB) catalog by \citet{liu06:HMXB}. We also
checked the Low-Mass X-ray binary (LMXB) catalog by
\citet{liu07:LMXB}, the INTEGRAL source catalog \cite{INTEGRAL:3IBIS},
and the SIMBAD database. Sky maps for \UPDATED{longer}-wavelength survey data in
the radio and X-ray wavebands, from the Molonglo
\cite{Molonglo,SUMSS}, NRAO VLA \cite{NVSS}, ROSAT \cite{ROSAT}, ASCA
\cite{ASCA} Galactic plane surveys, were compared with the HESS excess
maps. Additionally, pointed observations made by the Chandra and
XMM-Newton instruments were checked when available in the respective
archives. Unless otherwise noted, ROSAT survey data between 1.0--2.4
keV and ASCA data between 2--10 keV have been used.

To reduce the number of chance coincidences with cataloged sources,
some loose selection criteria were applied:

\begin{itemize} 

\item Based on previous detections in the VHE energy range
  \cite[e.g.][]{HESS:VelaX,HESS:RXJ1713}, we consider the association
  of a VHE source with a shell-type SNR plausible only if the VHE
  emission roughly matches the angular size of the remnant and is not
  significantly offset.

\item Due to the large number of cataloged pulsars in the galactic
  plane, only those which are energetic enough to power a PWN which
  could produce VHE emission were considered. A useful quantity for
  determining the possibility of VHE emission from pulsars is the
  spin-down flux measured at the solar-system, $\dot E/D^2$, where
  $\dot E$ is the spin-down luminosity and $D^2$ is the distance to
  the object (both measurable quantities)
  \cite{fierro95:_egret_pulsar_studies}. Defining the conversion
  efficiency, $\eta$, as the ratio of the integral energy flux of a
  gamma-ray source over a typical energy range (e.g. 200 GeV to 20
  TeV) to the pulsar spin-down flux at the solar system, we find that
  for typical spectral characteristics of the sources discussed here,
  $\dot E/D^2$ must be well above
  $10^{33}\:\mathrm{erg\:sec^{-1}\:kpc^{-2}}$ to produce the observed
  emission, even assuming 100\% efficiency; for this reason, pulsars
  with lower spin-down fluxes are not plotted in the figures given
  later in this paper. In cases where the distance estimate is not
  known, we assume a distance of 3 kpc. We note that efficiencies
  greater than 100\% are not completely excluded, since the spin-down
  flux might have been higher in the past and the particle cooling
  times might be comparable to the pulsar's age
  \cite{HESS:pwn_statistics}. However, to claim a plausible
  identification of a VHE source with a pulsar, we require
  efficiencies of $<10\%$ and a reasonably small angular distance for
  the purpose of this study to keep the number of chance coincidences
  low, unless there are other multi-frequency data that would support
  the association.
  
\item XRBs from which VHE emission is established are HMXBs that
  either exhibit a jet
  \cite[e.g. LS~5039,][]{HESS:LS5039_variability}, or where the
  compact object is a pulsar \UPDATED{powering} a PWN \cite[e.g. PSR
  B1259,][]{HESS:PSRB1259}; all appear variable and point-like in the
  VHE band. We believe the chance probability of the appearance of an
  XRB within the $3\sigma$ contours of an extended HESS source to
  be \UPDATED{reasonably} low, therefore we discuss such associations, but for
  the moment ignore XRBs lying outside the sources. Although the
  possibility exists that such an object might also power an extended,
  possibly asymmetric VHE source \cite[e.g.][]{cheng06:XRB_PWN}, this
  has so far not been observed.
  
\end{itemize}

\section{Results}


\begin{table}
  \begin{center}
    \begin{tabular}{l c@{$\:\pm\:$}c c@{$\:\pm\:$}c c@{$\:\pm\:$}c }
      \hline\hline
      Source    
      & \multicolumn{2}{c}{$\sigma_1$ ($^\circ$)} 
      & \multicolumn{2}{c}{$\sigma_2$ ($^\circ$)}  
      & \multicolumn{2}{c}{Angle ($^\circ$)} \\
      \hline
      \HESSa &  0.04  &  0.02  &  0.08   &  0.03  & 80 & 17 \\
      \HESSb &  0.07  &  0.02  &  0.10   &  0.05  &  3 & 40 \\
      \HESSc &  0.30  &  0.02  &  0.15   &  0.01  & 68 & 7\\
      \HESSd &  0.06  &  0.01  &  0.08   &  0.01  & -20 & 23\\
      \HESSe &  0.18  &  0.07  &  0.11   &  0.03  & -89 & 21\\
      \HESSf &  0.41  &  0.04  & 0.25    &  0.02  & 39 & 6\\
      \HESSg &  0.11  &  0.08  &  0.08   &  0.03  & -3 & 49\\
      \HESSh &  0.08  &  0.02  &  0.02   &  0.04  & 4 & 17\\
      \hline\hline
    \end{tabular}
    \caption{\label{tab:morphology} Results from an elongated 2-D
      Gaussian model fit (see \S\ref{sec:technique}) to the gamma-ray
      excess for each source. $\sigma_1$ and $\sigma_2$ are the
      intrinsic semi-major and semi-minor axes (in degrees on the
      sky), with the effect of the point-spread function removed. The
      errors are statistical. The position angle is measured
      counter-clockwise in degrees relative to the RA axis.}
  \end{center}
\end{table}  

\begin{table*}
  \begin{center}
    \begin{tabular}{l | c c c c  |r@{$\:\pm\:$}l  r@{$\:\pm\:$}l
        | c c c }
      \hline\hline
      Source 
      & $\theta_{int}$ 
      & Time 
      & $S$
      & Excess
      & \multicolumn{2}{c}{$\Gamma$} 
      & \multicolumn{2}{c}{$N_0\cdot 10^{-12}$}
      & $E_{\mathrm{min}}$ 
      & $E_{\mathrm{max}}$ 
      & $\frac{\chi^2}{D.O.F.}$
      \\
      & ($^\circ$) 
      & (hrs)     
      & ($\sigma $)         
      & (counts)  
      &\multicolumn{2}{c}{ }   
      &\multicolumn{2}{c}{\tiny ($\mathrm{cm^{-2}s^{-1}TeV^{-1}}$)}  
      & (TeV)
      & (TeV)
      \\
      \hline
      \HESSa  &  0.2 & 21 &  5.6 & 165 & 2.16  & 0.14 & 1.3  & 0.4 & 0.97 & 50 & 4.6/6  \\
      \HESSb  &  0.5 & 8  &  6.0 & 167 & 2.18  & 0.12 & 4.9  & 0.9 & 0.60 & 50 & 6.4/7  \\
      \HESSc  &  0.6 & 7  & 10.6 & 596 & 2.07  & 0.08 & 9.1  & 1.1 & 0.50 & 50 & 11.3/7  \\
      \HESSd  &  0.3 & 45 & 10.3 & 542 & 2.46  & 0.08 & 2.7  & 0.3 & 0.50 & 60 & 5.4/5  \\
      \HESSe  &  0.6 & 11 &  8.3 & 495 & 2.26  & 0.10 & 6.1  & 0.8 & 0.50 & 80 & 2.8/6  \\
      \HESSf  &  0.7 & 10 & 10.7 & 723 & 2.41  & 0.08 & 12.8 & 1.3 & 0.54 & 80 & 10.4/6 \\
      \HESSg  &  0.46& 15 & 10.2 & 425 & 2.39  & 0.08 & 6.1  & 0.7 & 0.60 & 80 & 5.9/6  \\
      \HESSh  &  0.15& 23 &  7.4 & 116 & 2.17  & 0.12 & 0.6  & 0.1 & 0.50 & 80 & 4.9/6  \\     
      \hline\hline
    \end{tabular}
    \caption{\label{tab:spectra} Summary of spectral parameters for
      each source from a power-law fit to the spectral data
      ($E=N_0E^{-\Gamma}$) over the energy range
      $E_{\mathrm{min}}-E_{\mathrm{max}}$. The integration radius,
      $\theta_{int}$ is chosen to fully enclose each source.  Only
      data with observation positions offset less than $2^\circ$ from
      the source position were included. The errors shown are
      statistical; the systematic error is conservatively estimated to
      be 20\% on the flux and $\pm0.2$ on the spectral index. Plots of
      all spectra are given in Figure \ref{fig:spectra}.}
  \end{center}
\end{table*}

\begin{figure*}[h]
  \centering
  \resizebox{0.32\hsize}{!}{\includegraphics{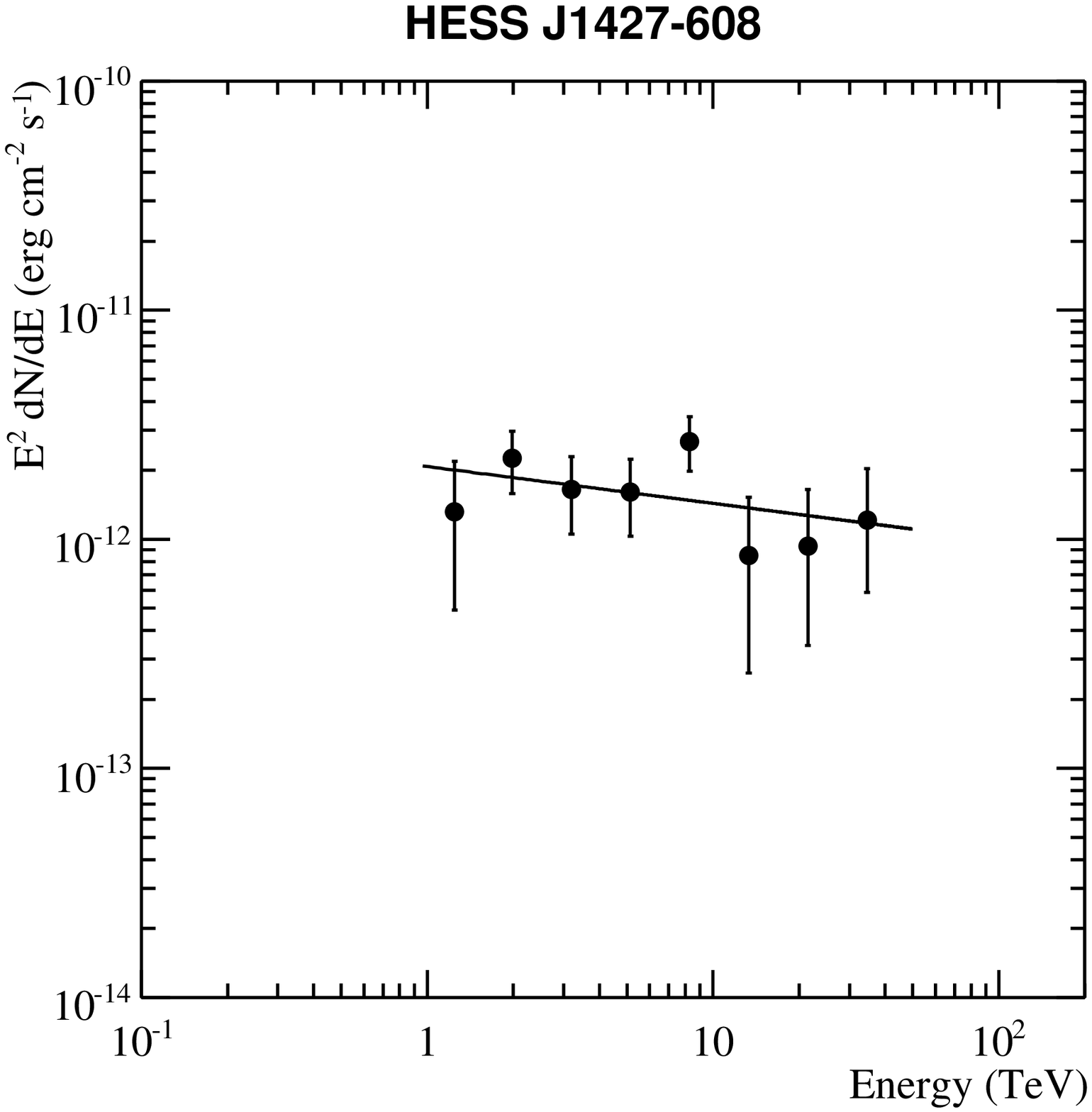}} 
  \resizebox{0.32\hsize}{!}{\includegraphics{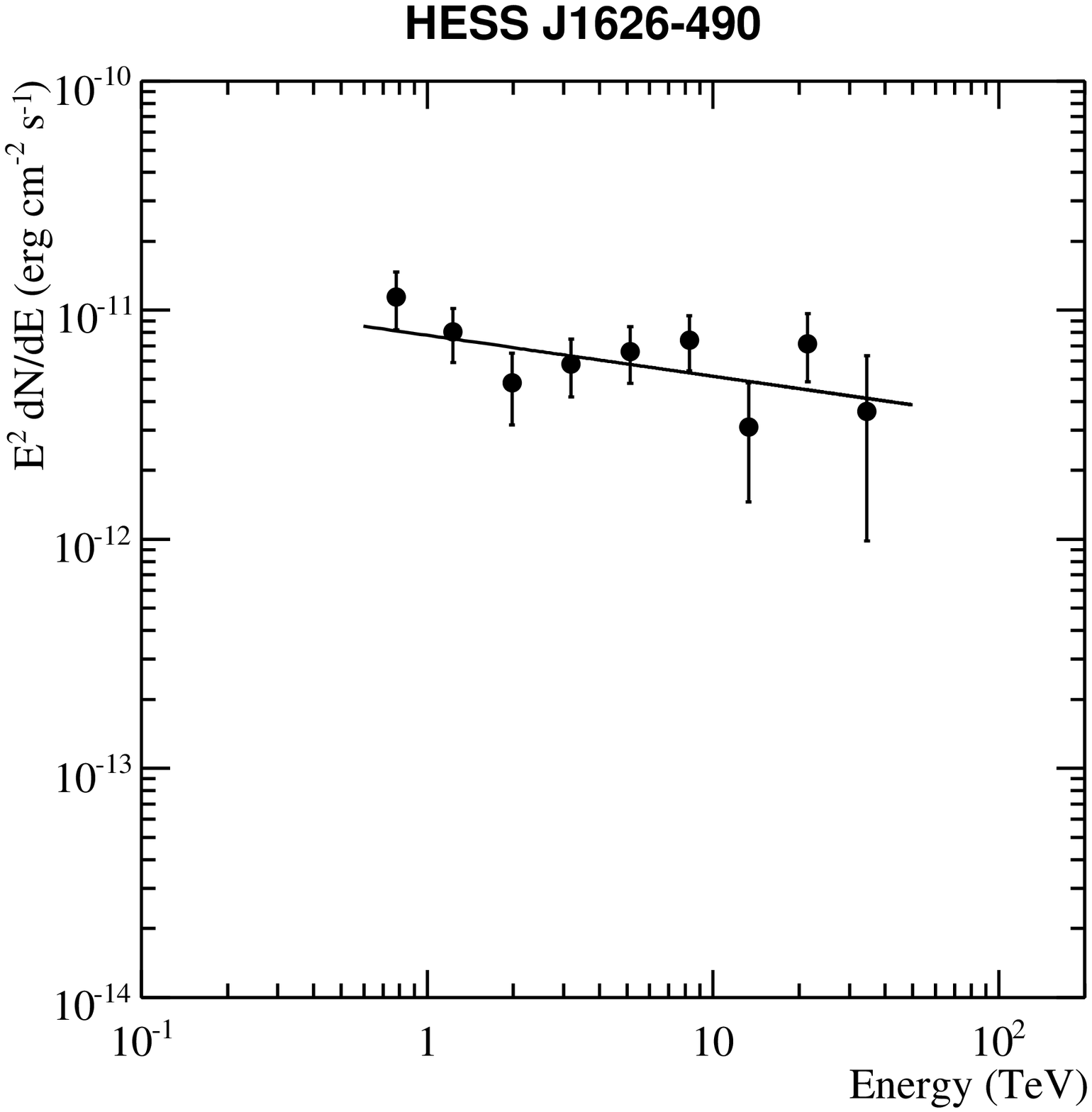}}\\
    \resizebox{0.32\hsize}{!}{\includegraphics{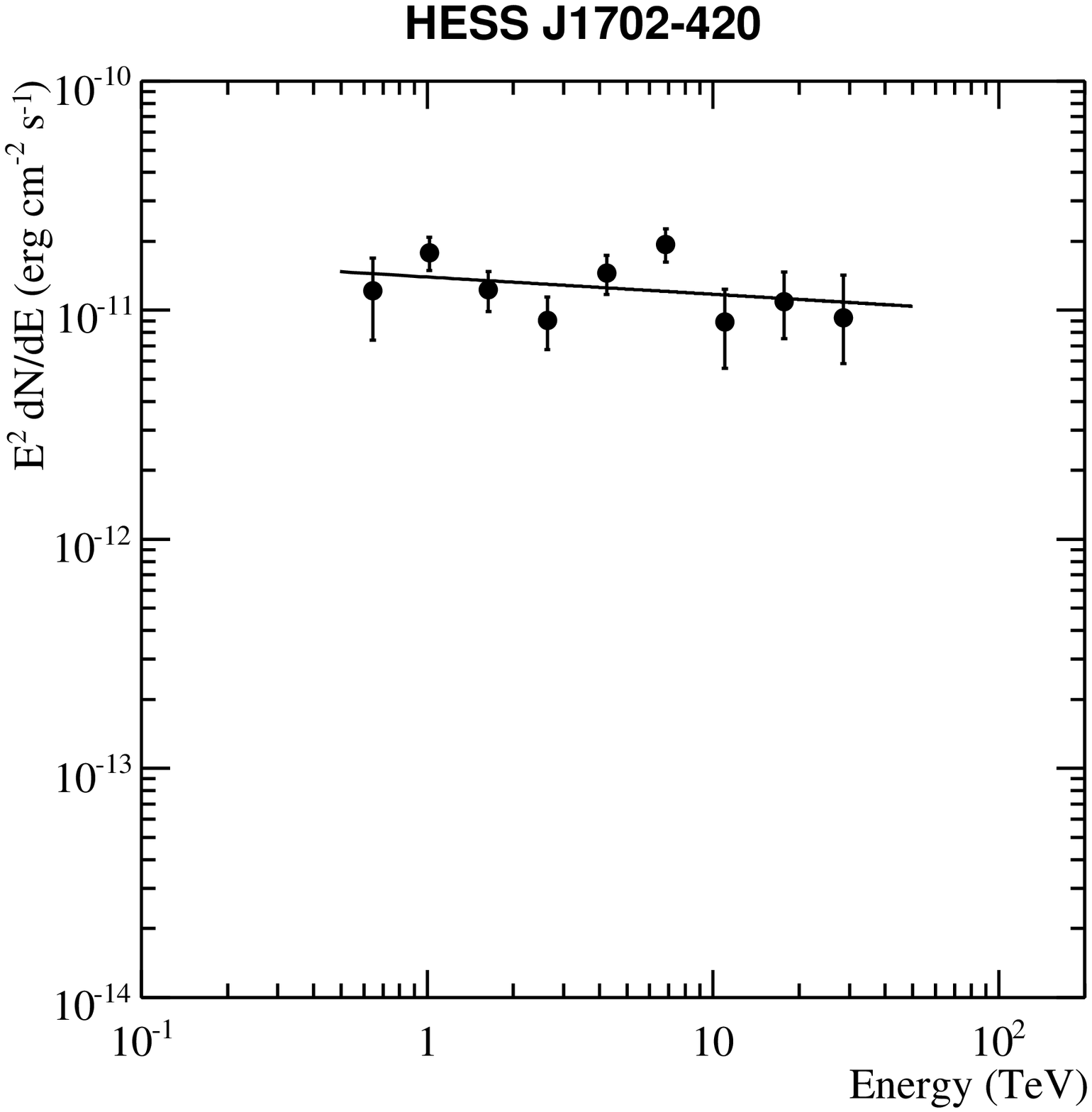}} 
    \resizebox{0.32\hsize}{!}{\includegraphics{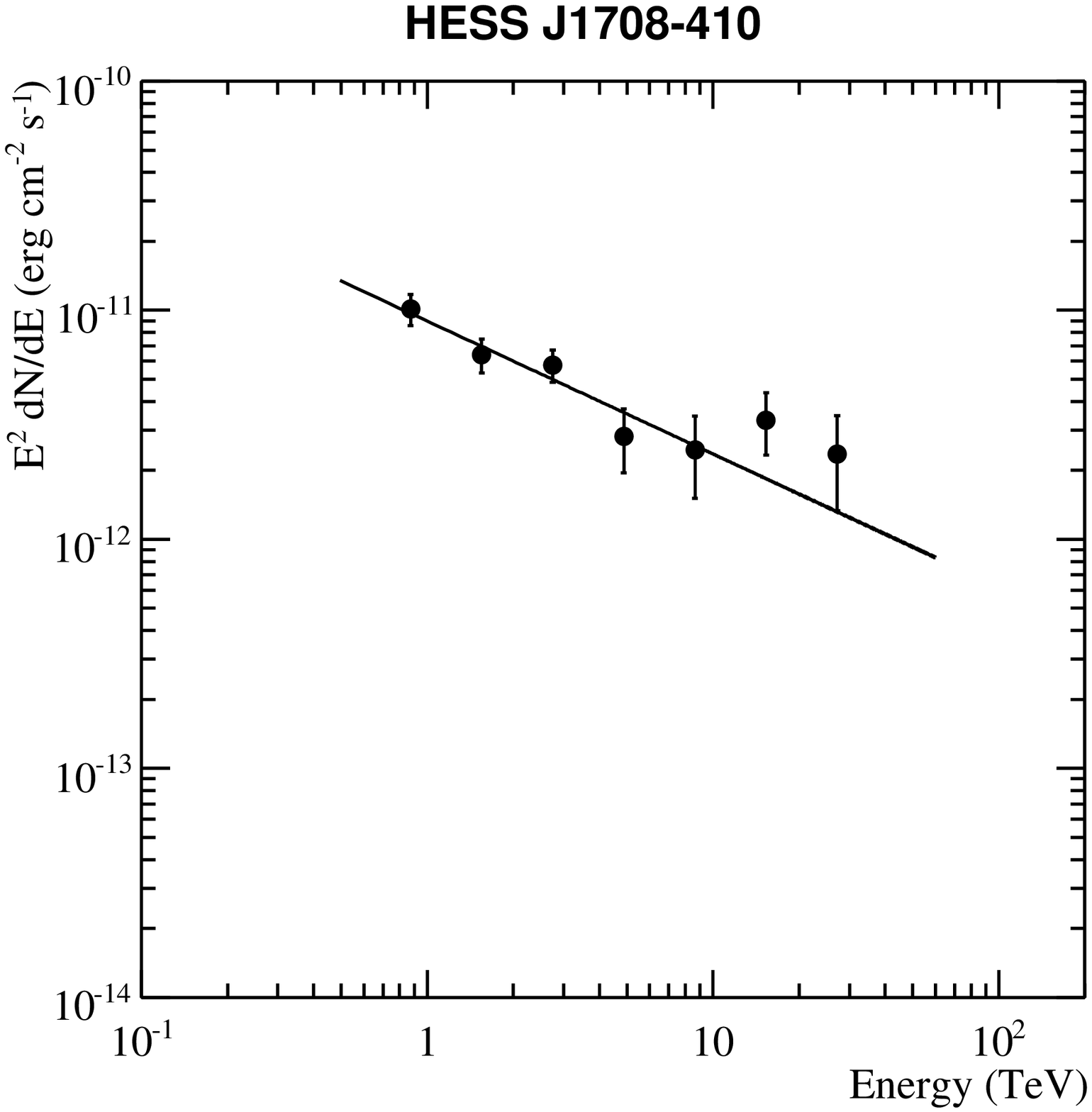}}\\
    \resizebox{0.32\hsize}{!}{\includegraphics{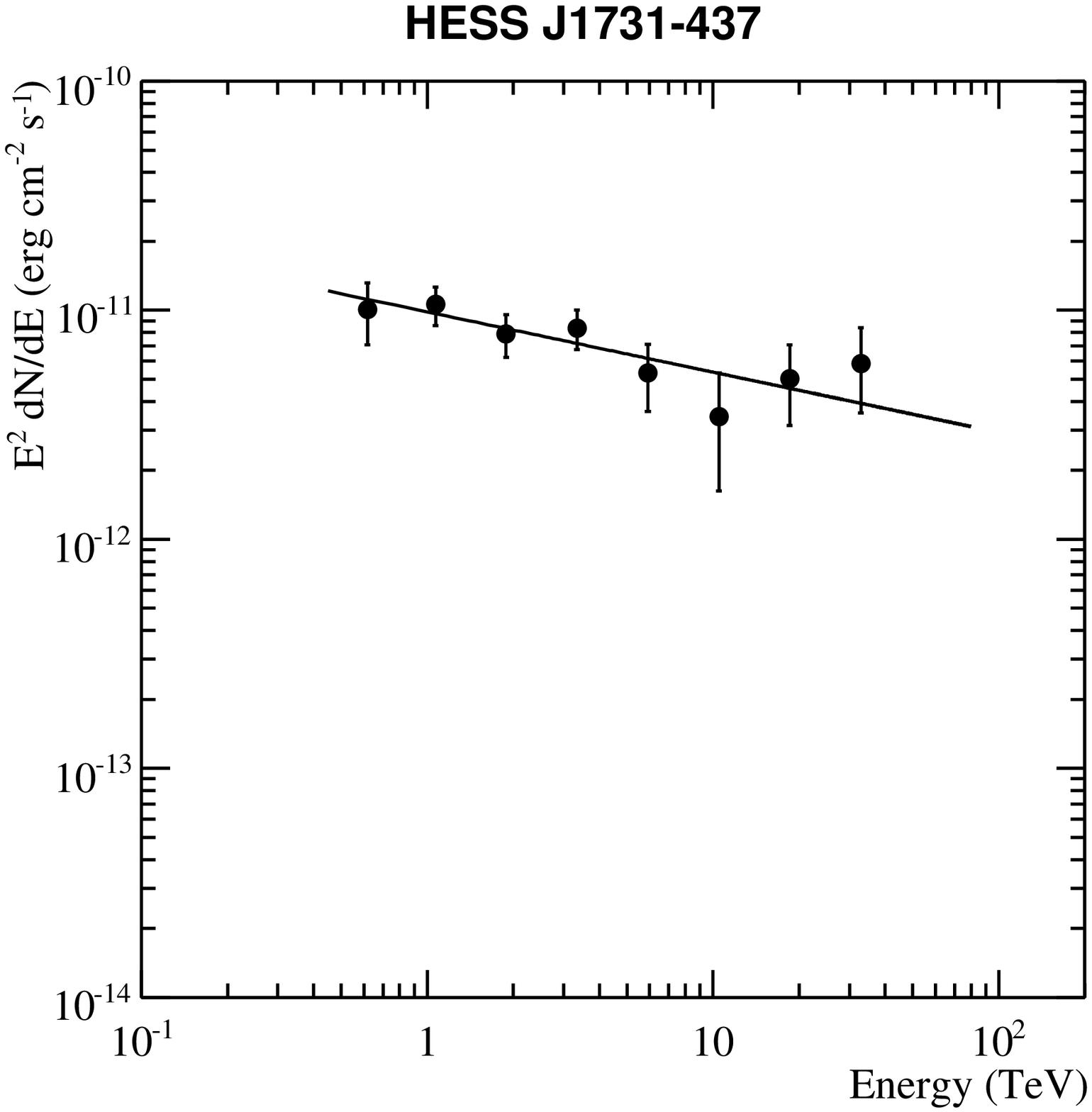}} 
    \resizebox{0.32\hsize}{!}{\includegraphics{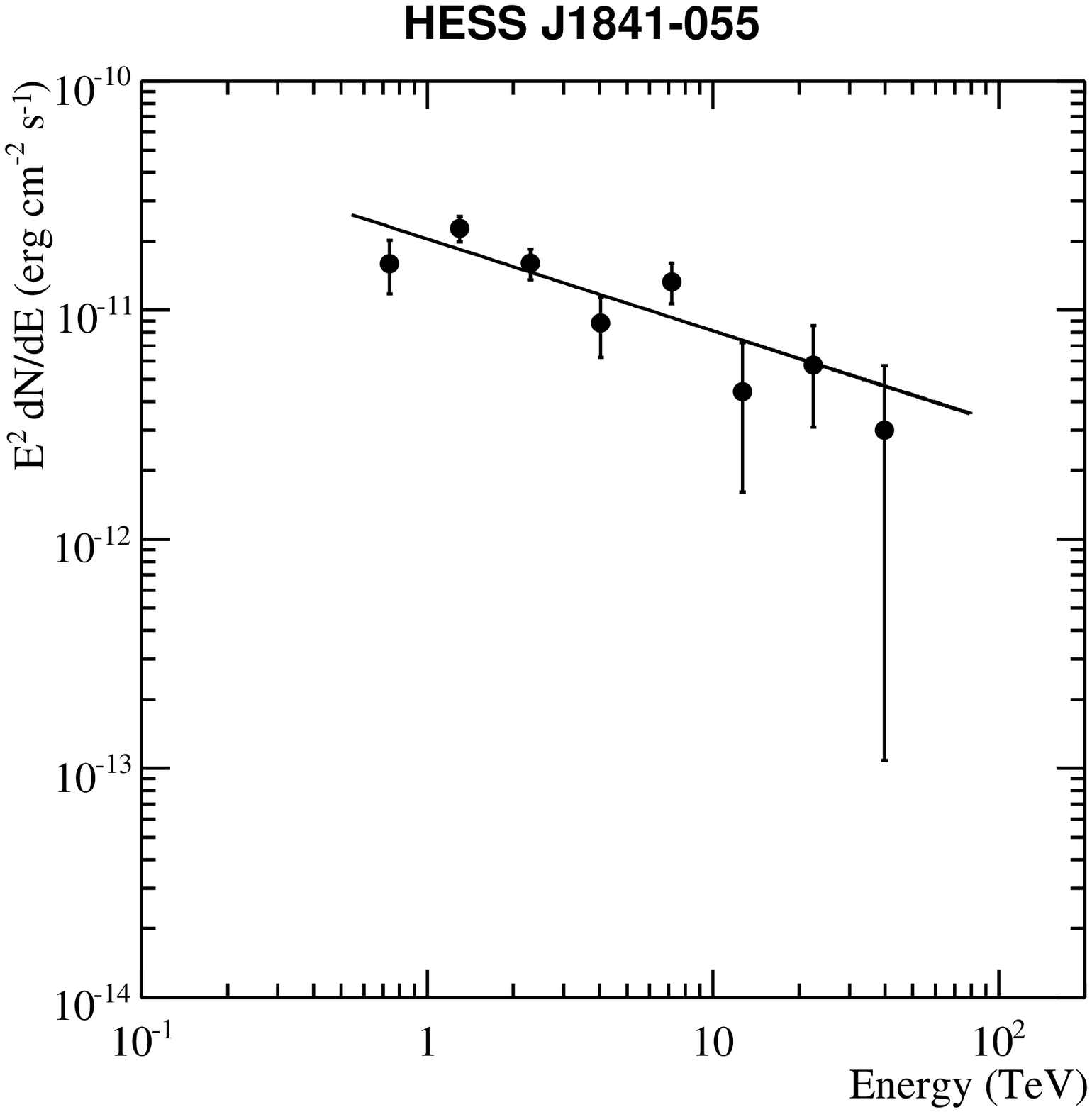}} \\
    \resizebox{0.32\hsize}{!}{\includegraphics{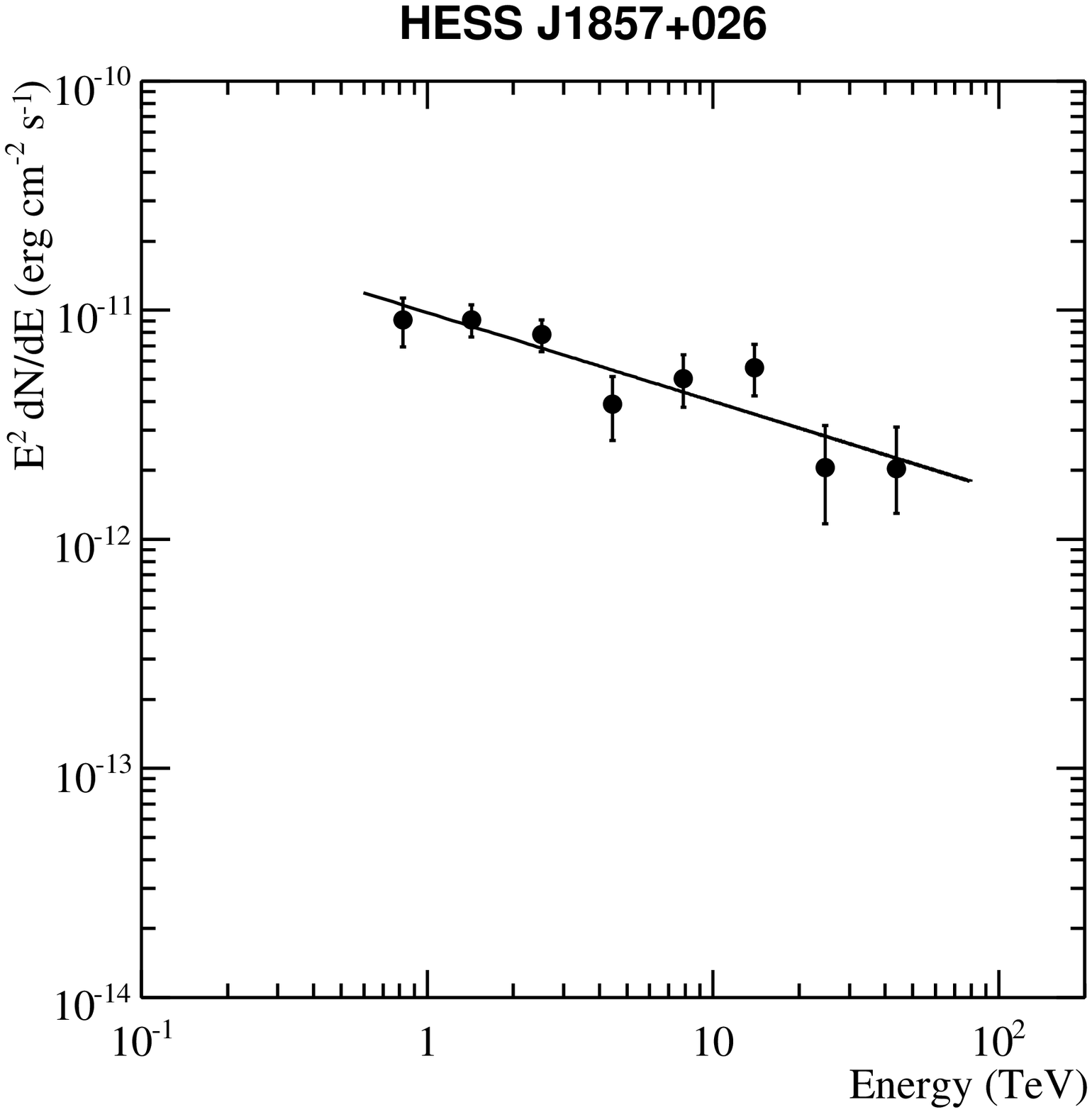}} 
    \resizebox{0.32\hsize}{!}{\includegraphics{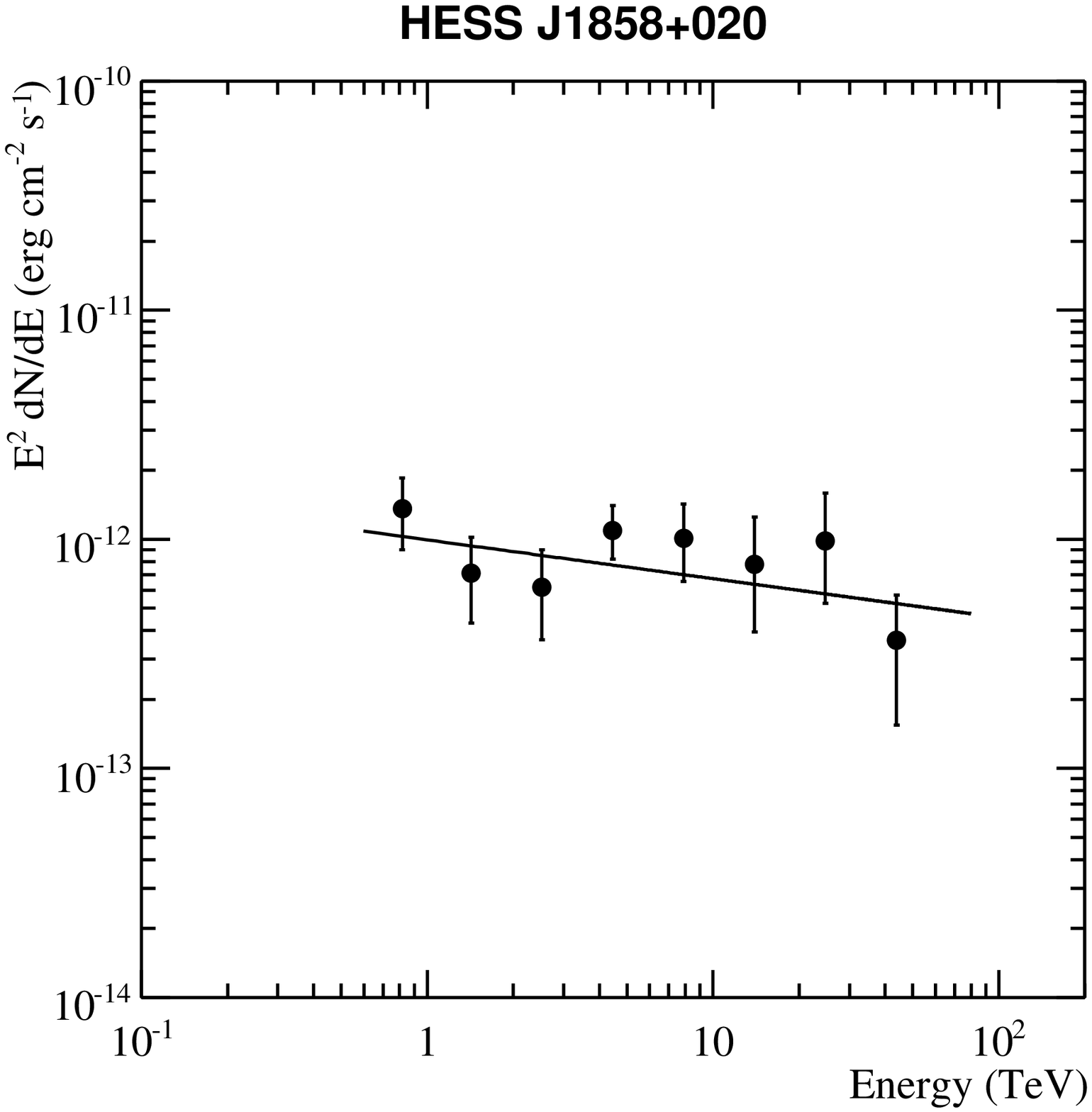}} 
    \caption{Spectra for each unidentified source, with power-law
      fits. See Table \ref{tab:spectra} for detailed fit
      information. }
    \label{fig:spectra}
\end{figure*}

Results of the size and spectral fits for each source are
summarized in Tables \ref{tab:morphology} and \ref{tab:spectra},
respectively. The spectrum for each source is plotted in Figure
\ref{fig:spectra}. In the following sections, a detailed
discussion of each source and related cataloged sources or
hot-spots within each field of view is given.

\subsection{\object{\HESSa}}

\begin{figure*}[h]
  \centering
    \resizebox{0.48\hsize}{!}{\includegraphics{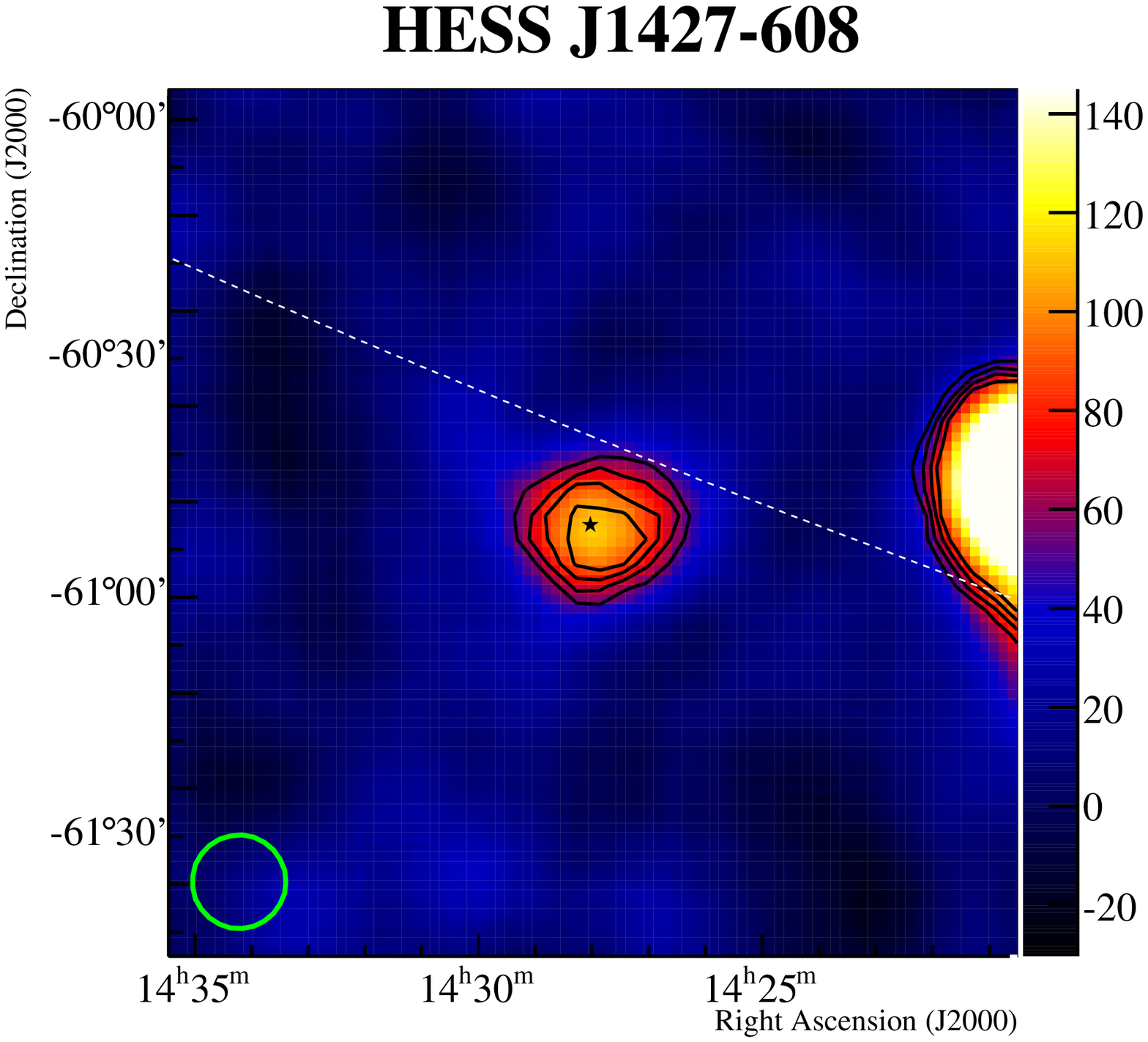}} 
    \resizebox{0.48\hsize}{!}{\includegraphics{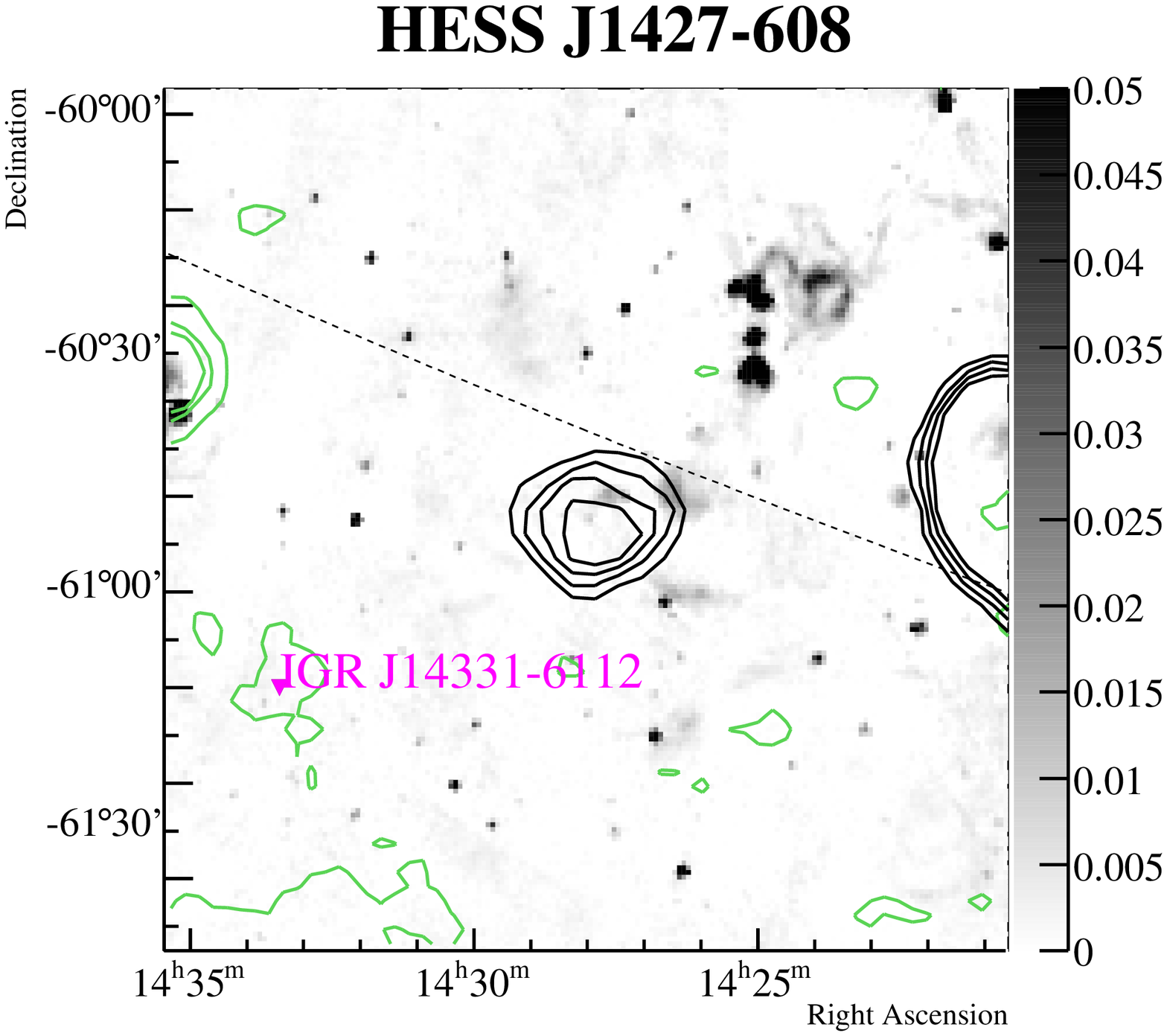}} 

    \caption{\emph{Left:} A VHE gamma-ray image of \HESSa\ (center
      position marked with a star). The image is of gamma-ray excess
      counts smoothed with a Gaussian filter with standard deviation
      $0.1^\circ$ (referred to as the \emph{smoothing radius}
      hereafter). The smoothing radius is chosen according to event
      statistics and therefore differs from source to source and is
      shown as a green circle in the lower left-hand corner.  The
      color scale of the image is set such that the blue/red
      transition occurs at approximately the $3\sigma$ (pre-trials)
      significance level.  Overlaid on the image are the significance
      contours starting at $4\sigma$ in $1\sigma$ steps. The Galactic
      plane is marked with a dashed line.  The gamma-ray excess at the
      right of the image is the hard X-ray source known as the
      Kookaburra/Rabbit, which is discussed in
      \citet{HESS:kookaburra}. \emph{Right:} The HESS
      significance contours (black) overlaid on a radio image
      \cite{Molonglo} (grey-scale, in Jy/beam). The green contours are
      from a ROSAT hard-band X-ray image \cite{ROSAT} which has been
      adaptively smoothed with the FTOOLS \emph{fadapt} algorithm to
      accentuate diffuse emission \cite{FTOOLS}. Also plotted are ATNF
      pulsars with $\dot E/D^2 \geq
      10^{33}\:\mathrm{erg\:s^{-1}\:kpc^2}$, SNRs from Green's
      catalog, HMXBs and LMXBs from the catalogs of Liu et al, and
      INTEGRAL sources (see Section \ref{sec:search} for references).
      In this case, only the INTEGRAL source IGR J14331-6112 lies
      within the field of view.  }
  \label{fig:J1427}
\end{figure*}

\HESSa\ (Figure \ref{fig:J1427}) is located approximately $1^\circ$
away from the hard X-ray and GeV gamma-ray source G313.2+0.3 (a strong
radio source located in the \emph{Kookaburra} complex) \cite{HESS:kookaburra},
and has a slightly extended morphology consistent with a symmetric
Gaussian of radius $\sigma=3'$. Its spectrum is fit by a power-law
with index \StatSysErr{2.2}{0.1}{0.2}. Radio and X-ray survey data of
the region (overlaid in Figure \ref{fig:J1427} from the Molonglo and
ROSAT surveys, respectively) show no evidence for significant emission
at distances of $0.5^\circ$ or closer to the centroid position of
\HESSa.  There are no nearby pulsars or supernova remnants, and an
association of \HESSa\ with the unidentified INTEGRAL source
IGR~J14331-6112 is unlikely due to the large angular distance
separating the two sources.

\subsection{\object{\HESSb}}

\begin{figure*}[h]  \centering
  \resizebox{0.48\hsize}{!}{\includegraphics{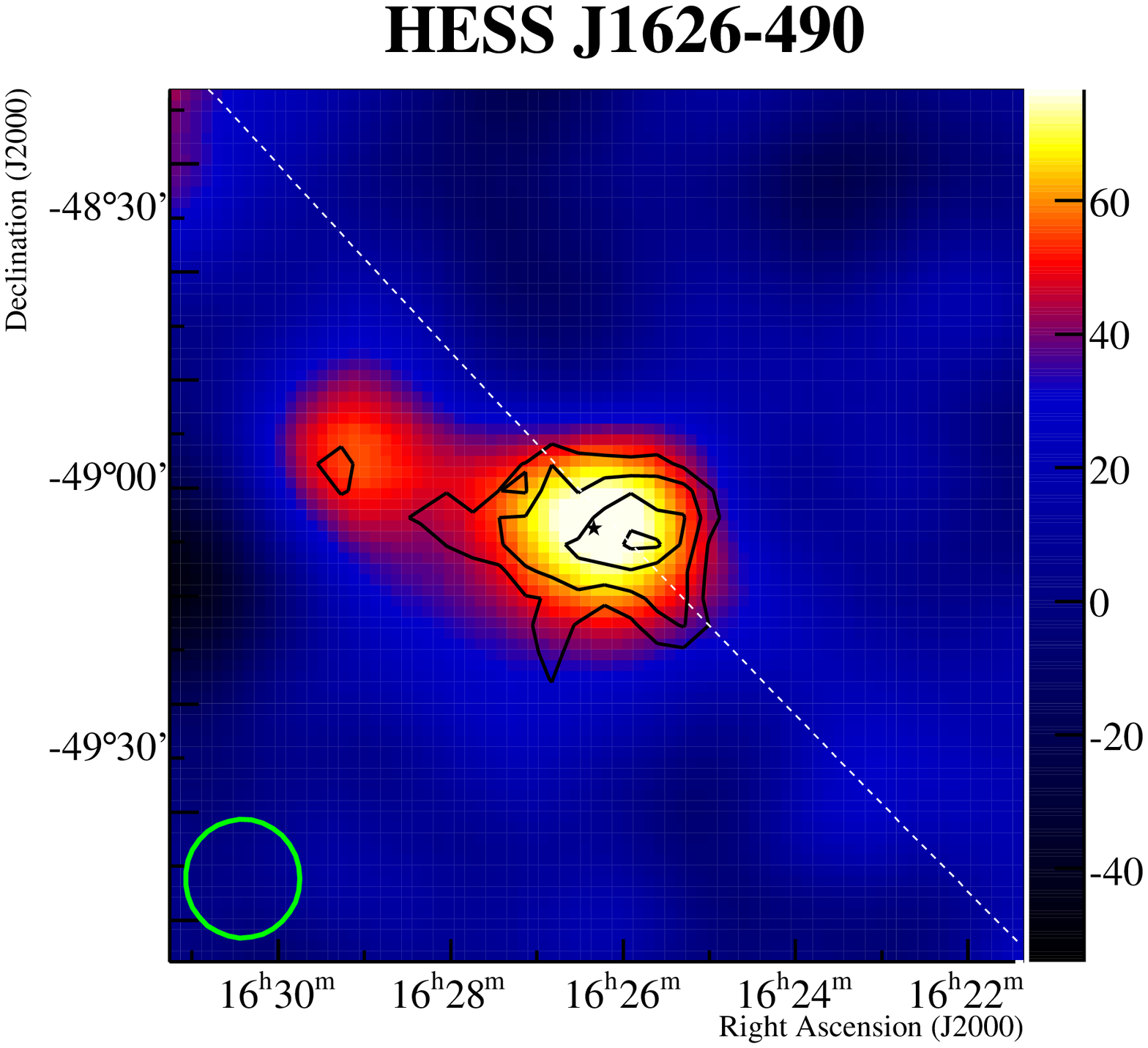}} 
  \resizebox{0.48\hsize}{!}{\includegraphics{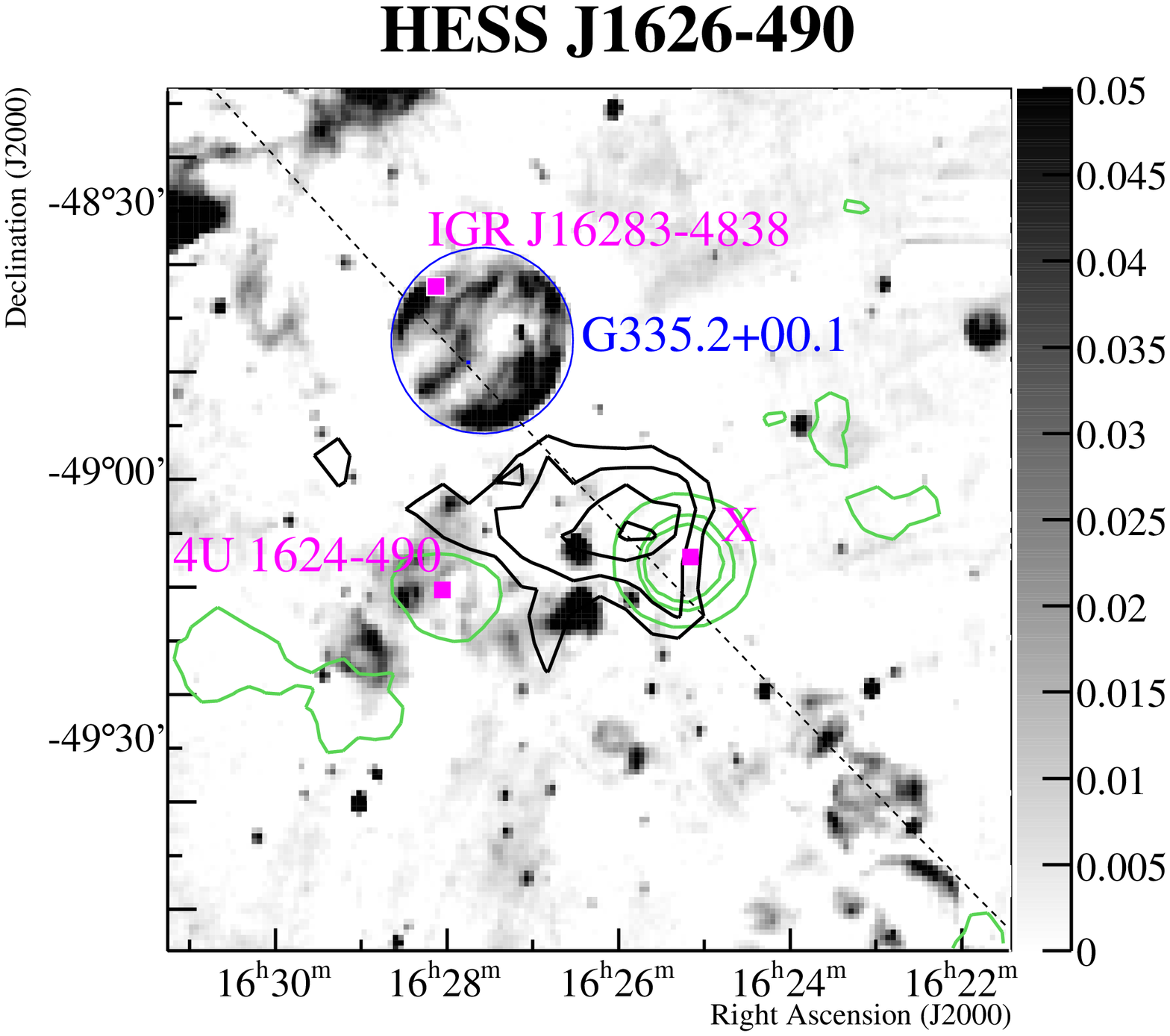}} 
  \caption{ Left: A VHE gamma-ray image of \HESSb\ plotted as in
    Figure \ref{fig:J1427}, with a smoothing radius of
    $0.1^\circ$. \emph{Right:} the HESS significance (black) and
    adaptively smoothed ROSAT X-ray contours (green), overlaid on the
    Molonglo radio image (grey-scale).  Also plotted is the SNR
    G335.2+00.1 (circle marking extent), the HMXB IGR~16283-4838, the
    LMXB 4U~1624-490, and the unidentified X-ray source
    1RXS~J162504-490918 (labeled $X$). }
  \label{fig:J1626}
\end{figure*}

\HESSb, located exactly on the Galactic plane (Figure
\ref{fig:J1626}), is a gamma-ray source with an approximately
radially-symmetric Gaussian morphology (with $5'$ extent), and a
power-law energy spectrum with photon index
\StatSysErr{2.2}{0.1}{0.2}. There is a slight extension toward
increasing right ascension which is only marginally significant, but
may be an indication of a second VHE source.  Within the gamma-ray
emission region, there exists some weak radio emission, along with the
unidentified X-ray source 1RXS~J162504-490918, which lies
approximately $10'$ from the centroid position and is a possible X-ray
counterpart. This X-ray source, marked with an ``X'' in the figure,
has an extent of $13''$ and an absorption-corrected flux between
0.1--2.0 keV of $1.7\times 10^{-13}\:\mathrm{erg\:cm^{-2}\:s^{-1}}$,
assuming a photon index of 2.0 \cite{ROSAT,WEBPIMMS}. The shell-type
supernova remnant G335.2+00.1 (MSH~16-4\textit{4}) \cite{MOST:MSH1644}
lies just outside the significant emission region of \HESSb, as does
the LMXB 4U~1624-490 \cite{smale00:_timin_two_lmxbs}, and the HMXB
IGR~16283-4838 \cite{INTEGRAL:3IBIS}, which are not considered
plausible candidates due to their offsets.


\subsection{\object{\HESSc}}

\begin{figure*}[h]
  \centering
  \resizebox{0.48\hsize}{!}{\includegraphics{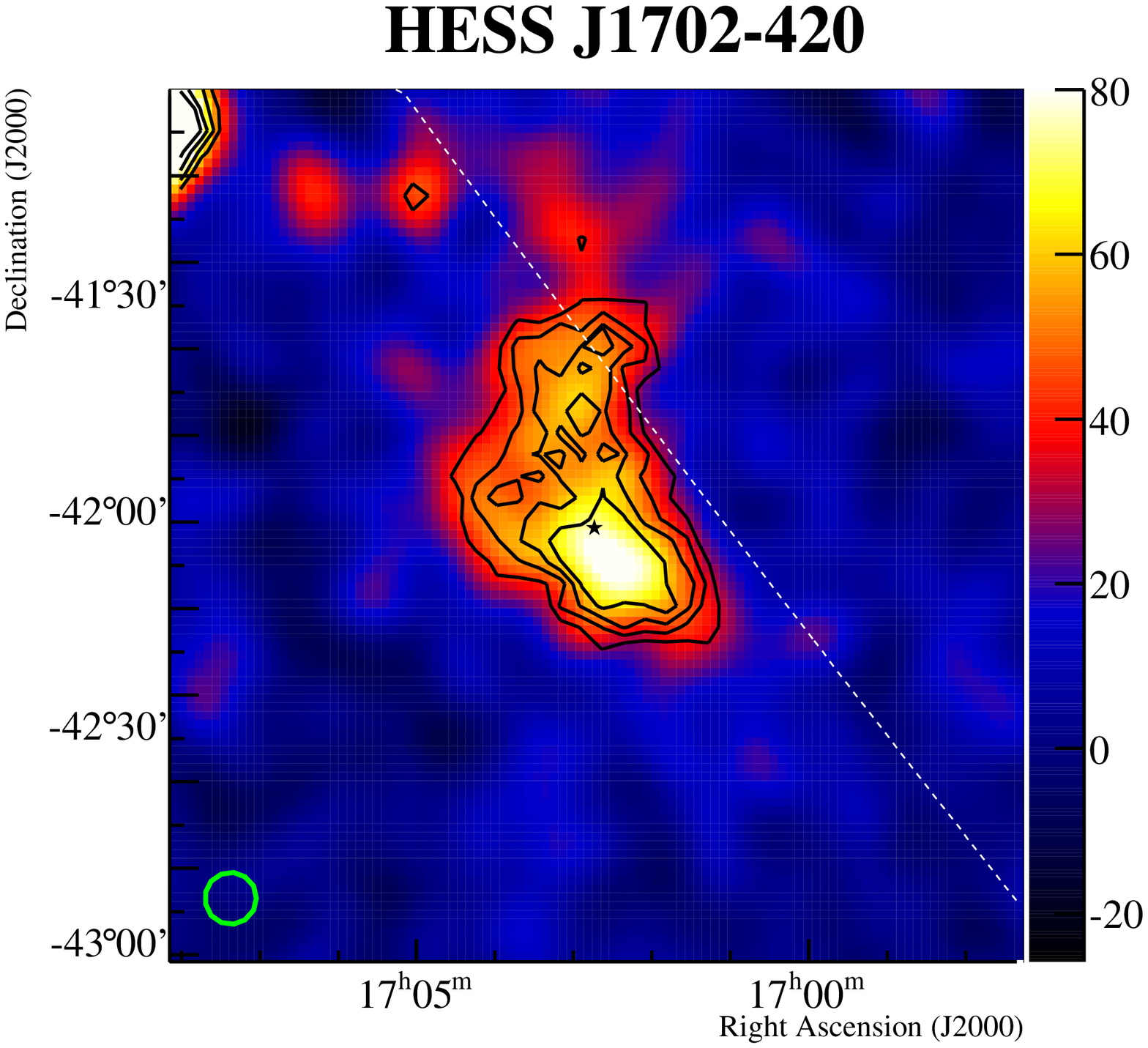}} 
  \resizebox{0.48\hsize}{!}{\includegraphics{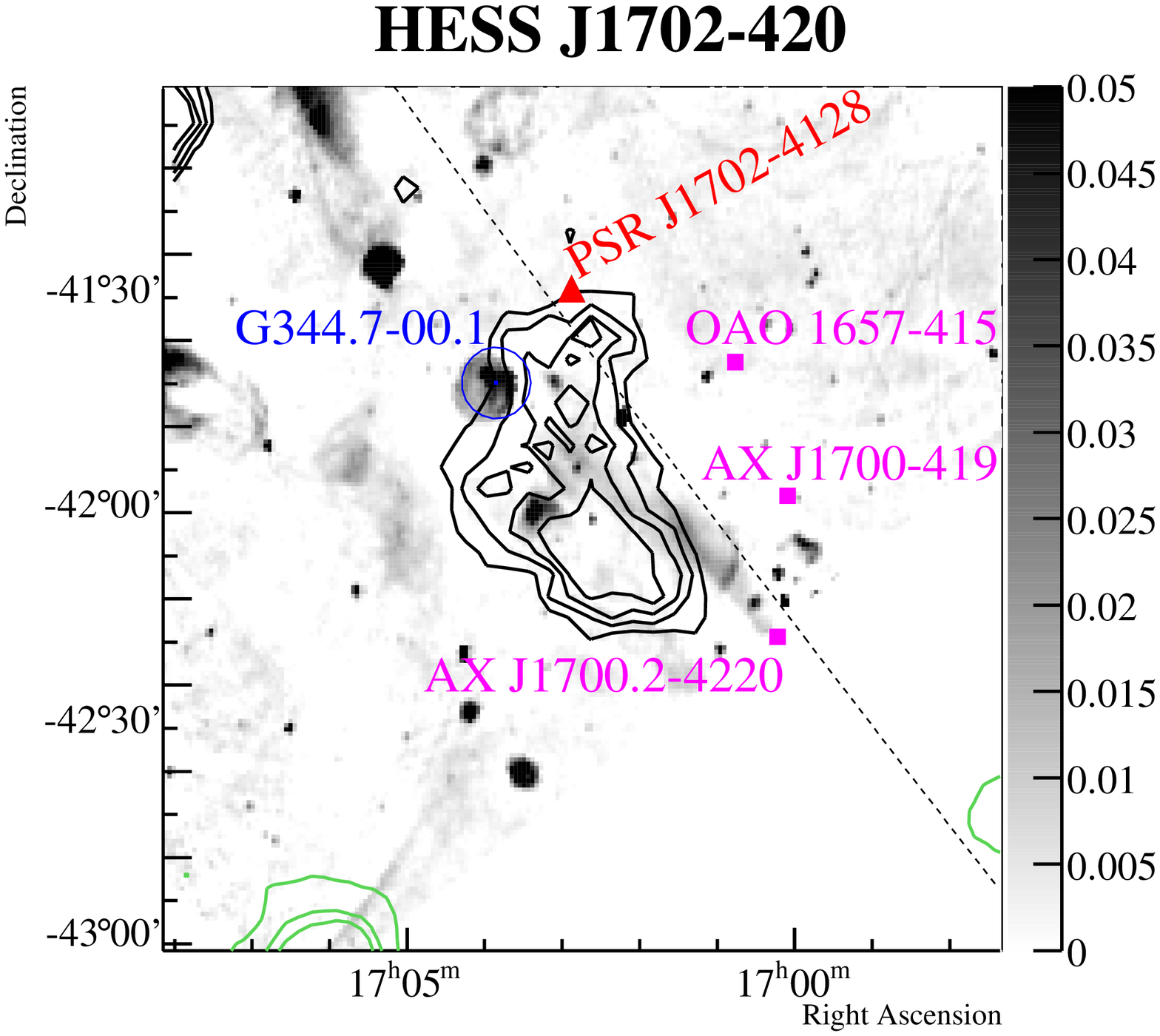}} 
  \caption{Left: A VHE gamma-ray image of \HESSc, plotted as in Figure
    \ref{fig:J1427}, with a  smoothing radius of
    $0.06^\circ$.  \emph{Right:} the HESS significance (black)
    and adaptively smoothed ROSAT X-ray contours (green), overlaid on
    the Molonglo radio image (grey-scale).  Also plotted are the
    positions of the SNR G344.7-00.1 (circle), three HMXBs
    (magenta squares), and the high spin-down flux pulsar PSR
    J1702-4128 (red triangle). }
  \label{fig:J1702}
\end{figure*}

First discovered by HESS at an approximately $6\sigma$
significance level \cite{HESS:scanpaper2}, \HESSc\ (Figure
\ref{fig:J1702}) is now seen with increased observation time at a
significance level of $13\sigma$.  Its spectrum is characterized by a
power-law with index \StatSysErr{2.1}{0.1}{0.2}, slightly harder than
the previously reported value of \StatSysErr{2.3}{0.15}{0.2}, which
was derived from a smaller integration radius, less statistics, and over
a smaller energy range. The results, including the source location,
are consistent within the errors.  The emission ``tail'' extending to
positive galactic longitude and latitude is statistically significant,
giving the source an elongated morphology (see Table
\ref{tab:morphology}).  The nearby pulsar {PSR J1702-4128} (to the
north of the VHE emission region, Figure \ref{fig:J1702}) lies at the
edge of the gamma-ray emission, and with $\dot
E/D^2=1.3\cdot10^{34}\:\mathrm{erg\:s^{-1}\:kpc^{-2}}$, it provides
enough spin-down energy loss to produce the observed emission
(assuming a rather high conversion efficiency of $\sim70\%$ if the
present distance estimate of 5 kpc is correct) and may be a
counterpart if it powers an extremely asymmetric pulsar wind nebula.
The nearby shell-type supernova remnant {G344.7-00.1} (seen in the
radio image) is also detected by ASCA in the 2--10 keV X-ray energy
band \cite{ASCA:_faint_x_ray_sourc_resol}, however is an unlikely
counterpart due to its small angular size and distance from the peak
of the emission region. Three X-ray binaries are also located nearby
the source (see the figure), but are outside the significant emission
region.

\subsection{\object{\HESSd}}

\begin{figure*}[h]
  \centering
  \resizebox{0.48\hsize}{!}{\includegraphics{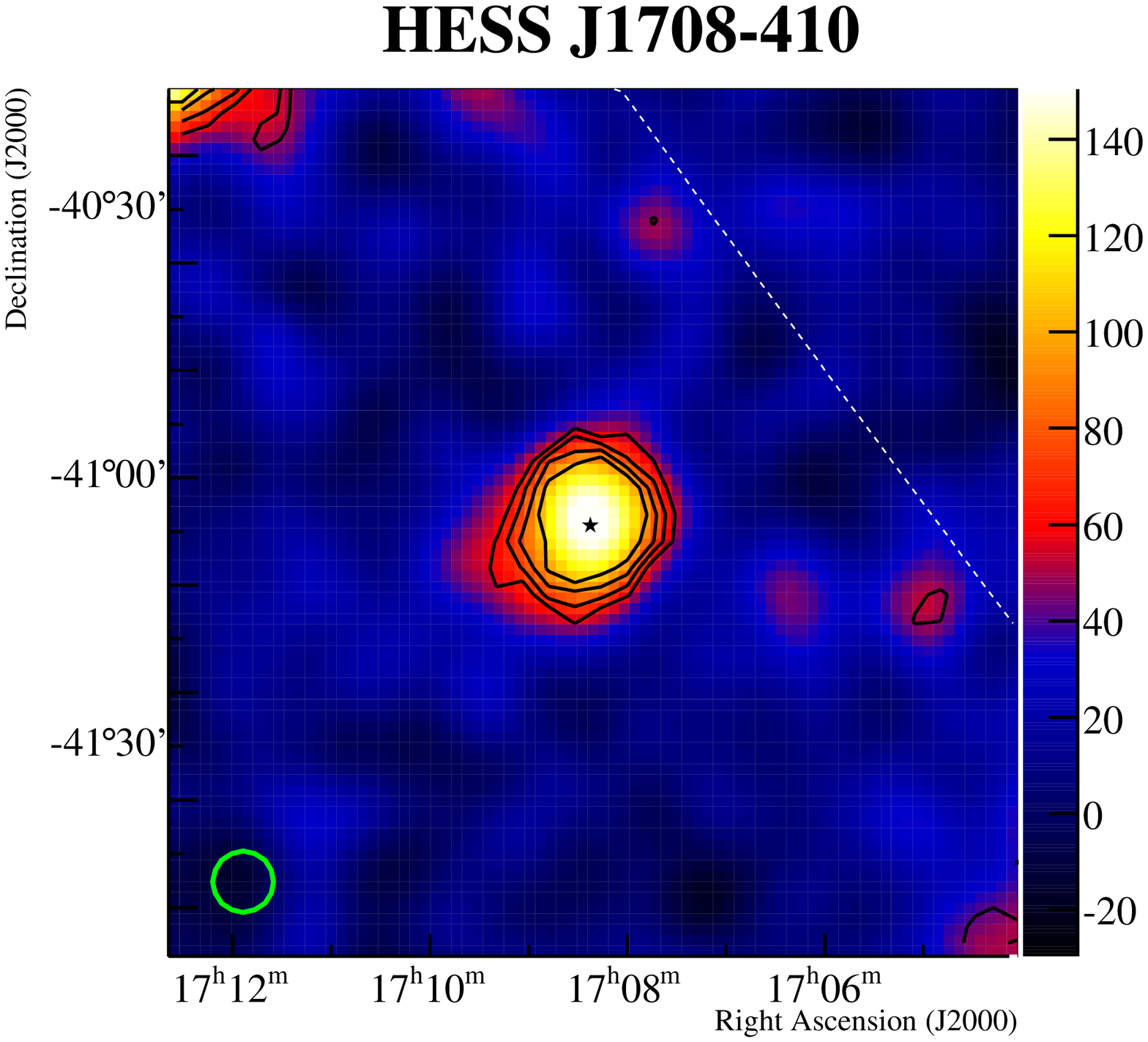}} 
  \resizebox{0.48\hsize}{!}{\includegraphics{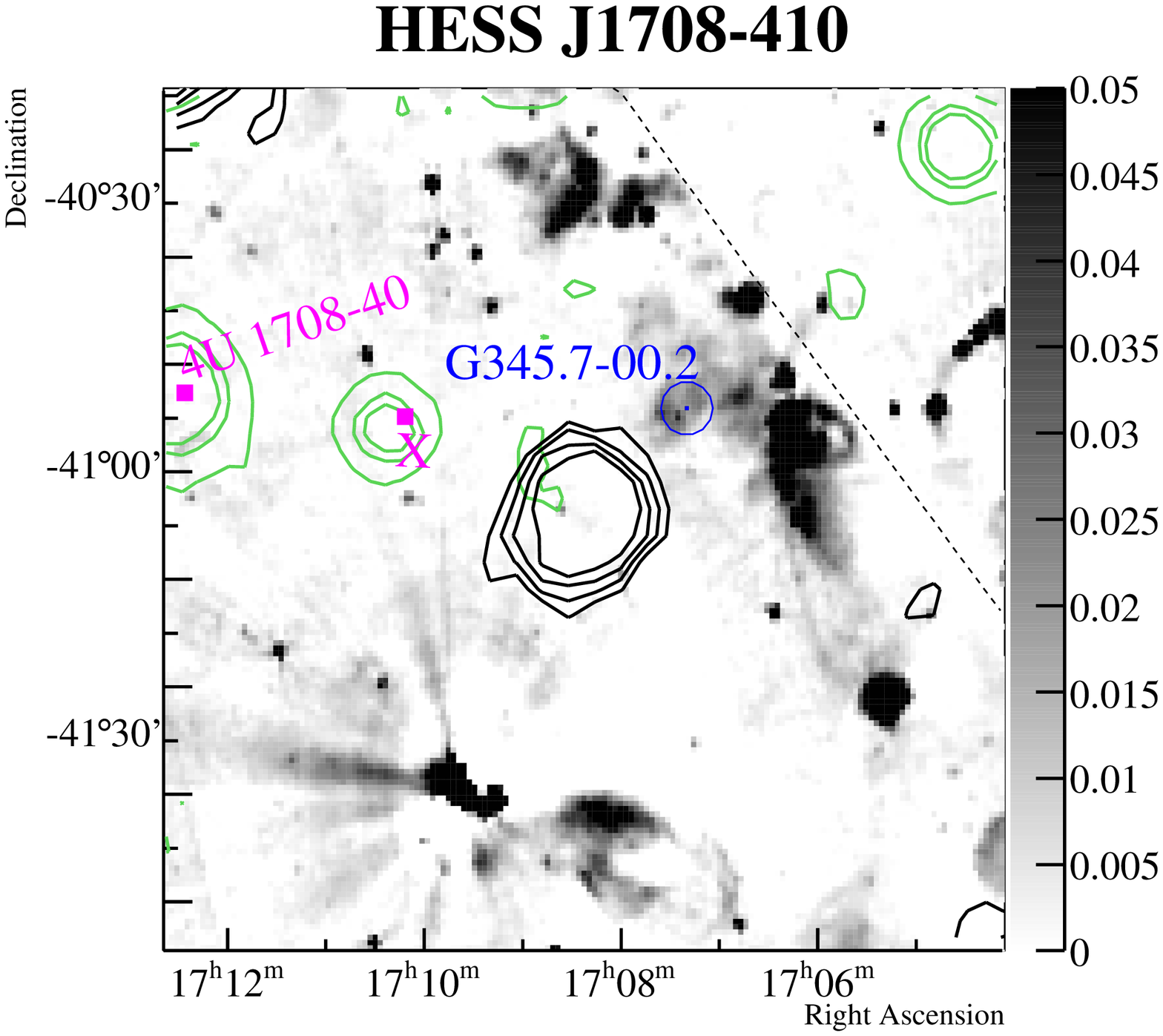}} 
  \caption{\emph{Left:} A VHE gamma-ray image of \HESSd\ plotted as in
    Figure \ref{fig:J1427}, with a  smoothing radius of
    $0.06^\circ$.  The slight excess seen on the lower-right
    corner of the image is \HESSc\ (see previous section), while the
    excess seen at the upper-left corner is part of RXJ~1713.7-3946
    \cite{HESS:RXJ1713}. \emph{Right:} The HESS significance
    (black) and adaptively smoothed ROSAT X-ray contours (green),
    overlaid on the Molonglo radio image (grey-scale).  Also plotted
    are the positions of the SNR G345.7-00.2, the ROSAT source
    1RXS~J171011.5-405356 (labeled ``X''), and the LMXB
    4U~1708-40 (square).  }
  \label{fig:J1708}
\end{figure*}

\HESSd\ (Figure \ref{fig:J1708}), situated between the supernova
remnant RXJ~1713.7-3946 \cite{HESS:RXJ1713} and \HESSc, was first
reported at a significance level of approximately $7\sigma$
\cite{HESS:scanpaper2}.  With additional observations of the region
(mostly from the edge of pointed observations centered on
RXJ1713.7-3946), the data set now has a statistical significance of
$11\sigma$.  The spectrum is fit by a power-law with index
\StatSysErr{2.5}{0.1}{0.2}, which is slightly softer than the
previously published result of \StatSysErr{2.3}{0.1}{0.2} made with
lower statistics, a smaller integration radius, and over a smaller
energy range \cite{HESS:scanpaper2}, though is within errors. The
compact morphology of \HESSd\ is consistent with a slightly elongated
Gaussian of approximately $0.08^\circ$ extent, with no significant
emission beyond $0.3^\circ$, ruling out SNR G345.7-00.2 or nearby
radio hot-spots as obvious counterpart candidates.  Although several
ROSAT hard-band X-ray hot spots exist in the field-of-view (e.g. the
XRB 4U~1708-40 or 1RXS~J171011.5-405356, see figure), the closest is
0.2$^\circ$ away and is not obviously connected with the gamma-ray
emission. There is an XMM-Newton exposure centered on {G345.7-00.2},
in which no significant emission is seen near the VHE
position. Additionally, an ASCA exposure of the region reveals only a
single point-like source located over a degree from the
HESS source.

\subsection{\object{\HESSe}}

\begin{figure*}[h]
  \centering
  \resizebox{0.48\hsize}{!}{\includegraphics{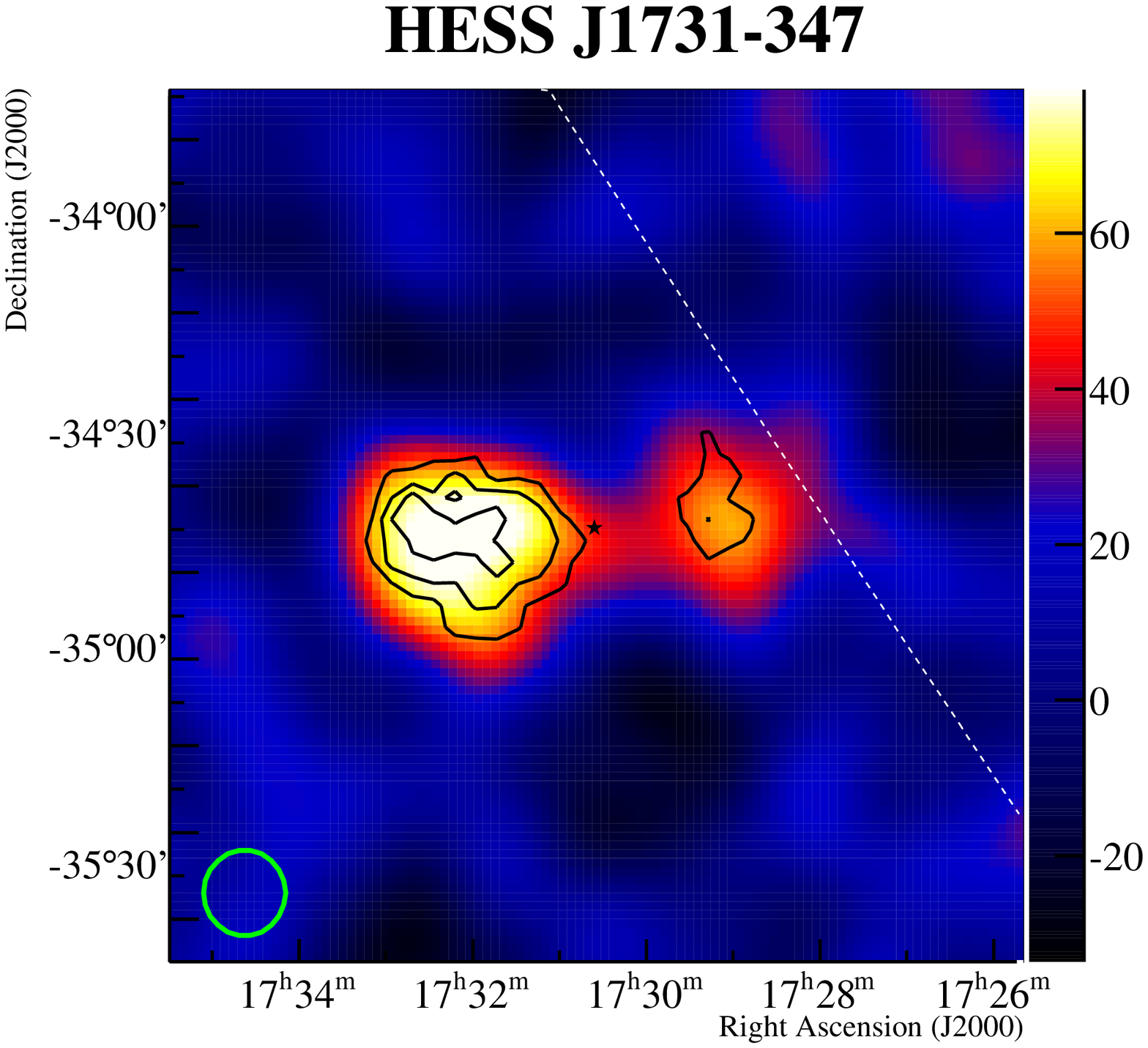}} 
  \resizebox{0.48\hsize}{!}{\includegraphics{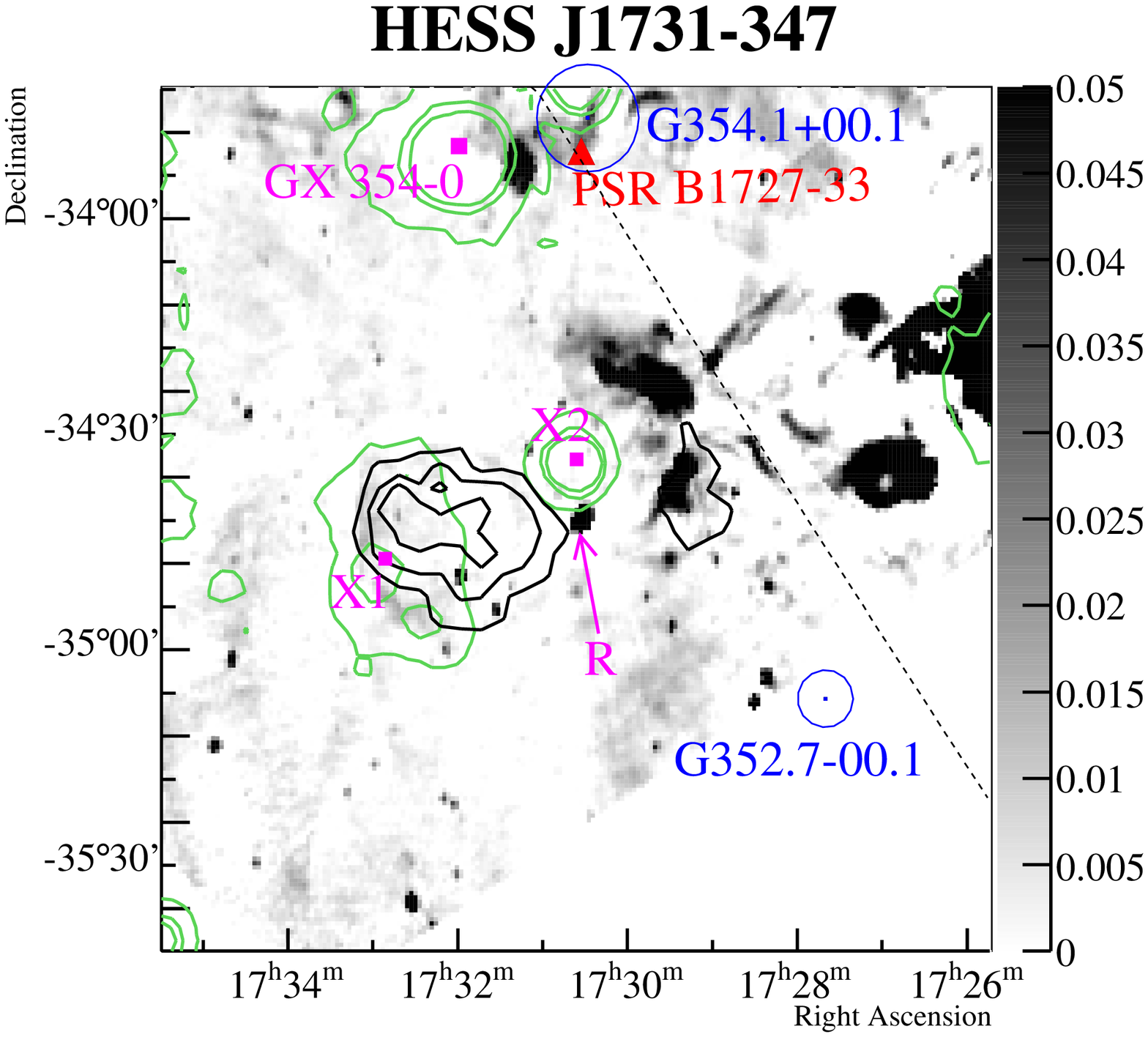}} 
  \caption{Left: A VHE gamma-ray image of \HESSe\, plotted as in
    Figure \ref{fig:J1427}, with a  smoothing radius of
    $0.1^\circ$. \emph{Right:} The HESS significance
    (black) and adaptively smoothed ROSAT X-ray contours (green),
    overlaid on the Molonglo radio image (grey-scale). Also shown are
    the positions of a high spin-down flux pulsar (filled
    triangle), the low-mass X-ray binary GX~354-0, and cataloged
    supernova remnants (blue circles marking extent). The source
    labeled $X_1$ is the ROSAT source 1RXS~J173251.1-344728, $R_2$ is
    the point-like radio source 173028-344144, and $X_2$ is the ROSAT
    point-source 1RXS~J173030.3-343219. }
  \label{fig:J1731}
\end{figure*}

\HESSe\ (Figure \ref{fig:J1731}) is detected at an $\sim{\hskip
  -0.3em}8\:\sigma$ level, exhibiting a power-law spectral index of
  \StatSysErr{2.3}{0.1}{0.2}. The source has a significant tail which
  extends westward, giving it a non-Gaussian morphology, possibly
  indicating the presence of more than one or an extended non-uniform
  source. A slice in the uncorrelated excess event map along the axis
  of the emission does not show a conclusive separation between the
  two ``peaks'', and a spectral analysis of each gives the same photon
  index within systematic errors. For this reason, the emission is
  treated here as a single source.

A bright X-ray point source ({1RXS~J173030.3-343219}, labeled as ``X''
in the figure) is seen in the ROSAT data, approximately 0.4 degrees in
the direction of the Galactic Plane from the centroid position, and
has an absorption-corrected flux in the 0.1--2.4 keV range of approximately
$2.0\times10^{-11}\:\mathrm{erg\:cm^{-2}\:sec^{-1}}$
\cite{ROSAT:1RXS,WEBPIMMS}, assuming a spectral index of 2.0. This source is
identified with the cataclysmic variable (CV) star HD~158394, and is
not expected to produce VHE emission. However, around the brightest part
of the TeV emission, there is some unidentified nebular X-ray emission
that partially matches the morphology of the HESS source, and may
well be the X-ray counterpart. This diffuse X-ray emission includes the
extended ROSAT source {1RXS~J173251.1-344728} (labeled $X_1$ in the
figure), which has an extension of $2'$ and X-ray flux of
$(7\pm1)\times10^{-12}\:\mathrm{erg\:cm^{-2}\:sec^{-1}}$, and a nearly
coincident point-like radio source labeled 353.464-0.69 in the VLA
survey data \cite{zoonematkermani90:VLA_radio}; their association
with the VHE emission is unclear. The strong point-like radio source
173028-344144 \cite{NVSS}, labeled $R$ in the figure, also lies to
the right of the peak of the VHE emission.  The X-ray emission about a
degree away to the north in the figure comes from the LMXB
GX~354-0, however due to its distance from \HESSe\ and since these
objects are not known to produce extended gamma-ray emission, it is an
unlikely counterpart candidate. No known high spin-down flux pulsars lie within
the emission region.

\subsection{\object{\HESSf}}

\begin{figure*}[h]  
  \centering
  \resizebox{0.48\hsize}{!}{\includegraphics{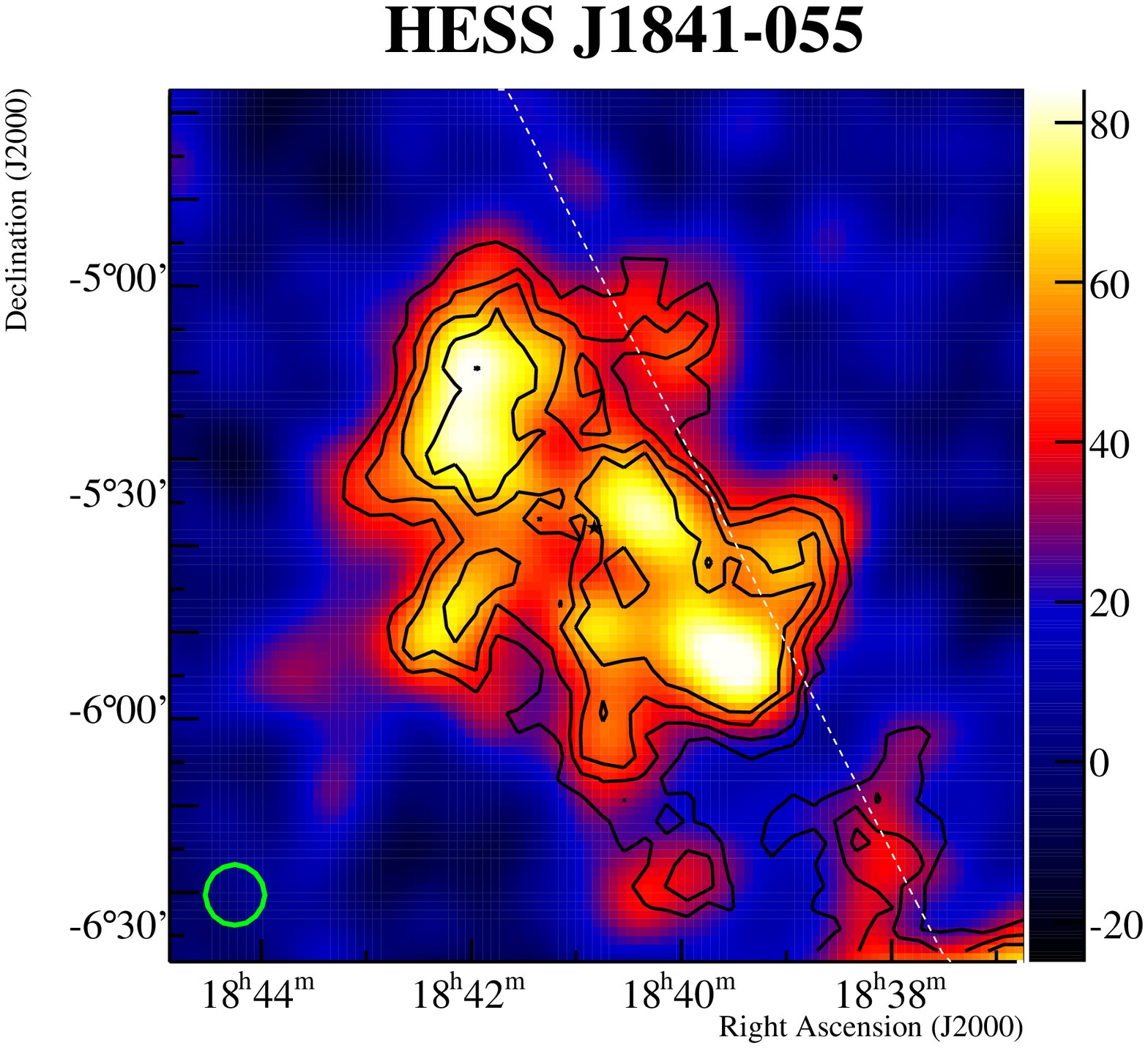}} 
  \resizebox{0.48\hsize}{!}{\includegraphics{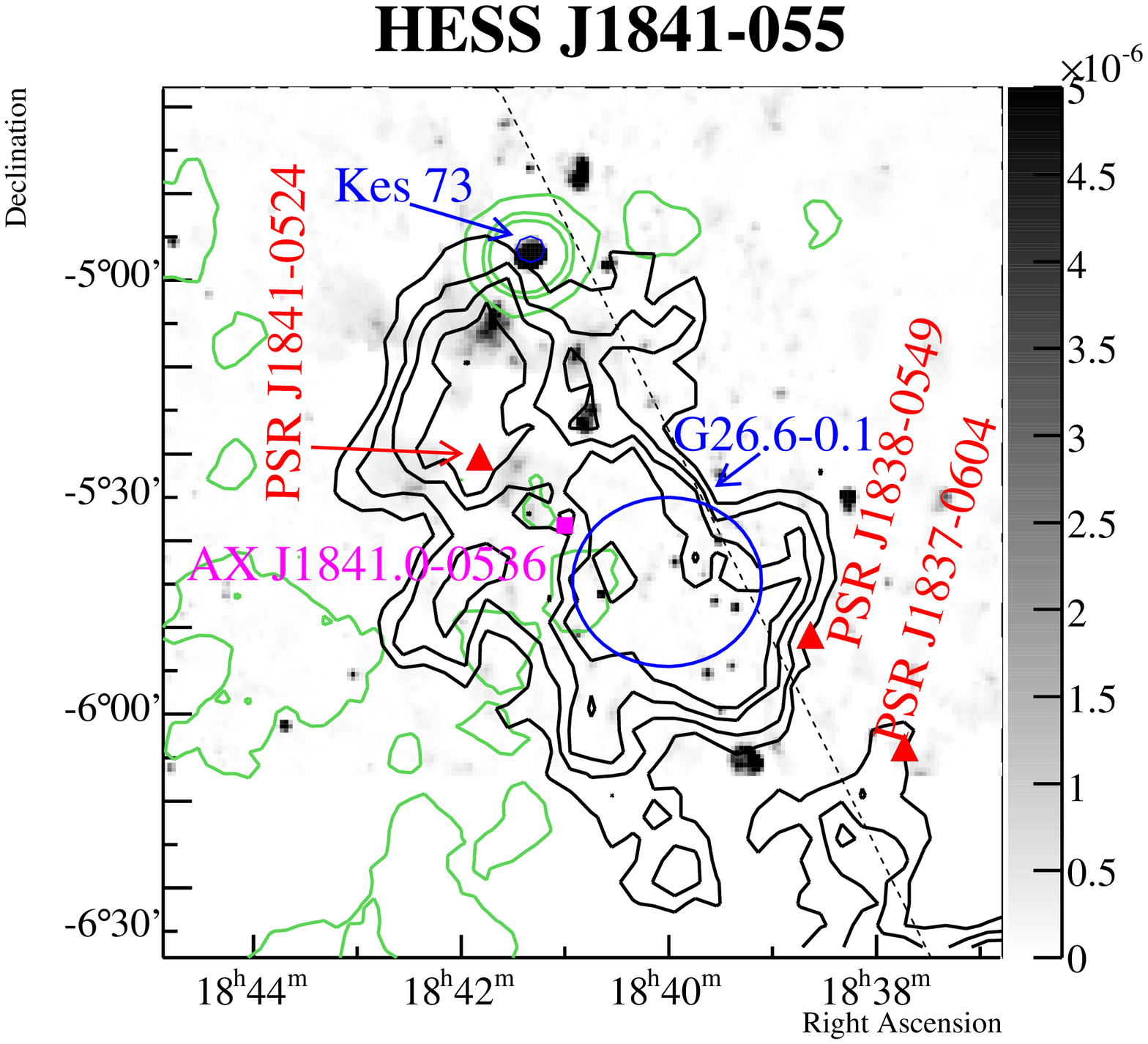}} 
  \caption{Left: A VHE gamma-ray image of \HESSf, plotted as in Figure
    \ref{fig:J1427}, with a  smoothing radius of
    $0.07^\circ$. \emph{Right:} The HESS significance (black)
    and adaptively smoothed ROSAT X-ray contours (green), overlaid on
    the NVSS radio image (grey-scale). Also shown are the
    positions of known high spin-down flux pulsars (filled triangles),
    the SNR Kes 73 (circle), the X-ray candidate SNR G26.6-0.1 (blue
    circle), and the HMXB AXJ~1841.0-0536. The Ginga source GS~1839-06
    is compatible with the location of AXJ~1841.0-0536.  }
  \label{fig:J1841}
\end{figure*}

\HESSf\ (Figure \ref{fig:J1841}) exhibits a highly extended, possibly
two or three-peaked , morphology; however, the ``dip'' between the
peaks along the major axis is not statistically significant
($<1.5\sigma$). The source has a spectrum that is fit by a power law
with index \StatSysErr{2.4}{0.1}{0.2}. An association with either
pulsar {PSR~J1841-0524} ($\dot E/D^2 =
4.4\cdot10^{33}\mathrm{erg\:s^{-1}\:kpc^{-2}}$) or PSR~J1838-0549
($\dot E/D^2 = 4.7\cdot10^{33}\mathrm{erg\:s^{-1}\:kpc^{-2}}$), is not
ruled out, however taken separately, each would require approximately
200\% efficiency to explain the VHE emission. This is not completely
implausible if both pulsars contribute together or if either had a
much higher spin-down luminosity in the past (PSR~J1838-0549 is
estimated to have a relatively old characteristic age of 112 kyr,
while {PSR~J1841-0524} is about 30 kyr old
\cite{ATNF}). PSR~J1837-0604 ($\dot
E/D^2=5.2\cdot10^{34}\mathrm{erg\:s^{-1}\:kpc^{-2}}$) has a high
enough spin-down flux to be a counterpart candidate, however it is
well outside the emission region. There are no cataloged PWN at \UPDATED{longer}
wavelengths identified with any of the three pulsars
\cite[e.g.][]{gotthelf04:_spin_power_thres_pulsar}. The SNR G027.4+00.0
(also known as {Kes 73}), which is visible in both X-ray and radio
wave bands, lies at the edge of the emission, though does not appear
related due to its small angular size.  Additionally, the high-mass
X-ray binary J1839-06 also lies near the edge of the significant TeV
excess.

ASCA observations of the Scutum arm region reveal a point-like source,
AX~J1841.0-0536, near the center of the VHE emission, which based on
its X-ray light curve and optical emission is suggested to be a
Be/X-ray binary pulsar \cite{bamba01:_discov_trans_x_ray_pulsar_ax_j1841}
with a flux in the 6--20 keV energy range of $1.1 \times 10^{-10}
\mathrm{erg\:cm^{-2}\:s^{-1}}$ and photon index of $2.2 \pm 0.3$
\cite{filippova05:INTEGRAL_1841}. A Chandra observation of this object
confirms the identification, with a flux in the 0.5--10 keV energy
range of $4.2\times10^{-12} \mathrm{erg\:cm^{-2}\:s^{-1}}$
\cite{halpern04:_chand_optic_ident_ax_j1841}. Given its point-like
extent, AX~J1841.0-0536 is not large enough to explain the entire
HESS  source, however it may well be responsible for a component
of the emission. 

Also within the VHE emission region lies the diffuse source
{G26.6-0.1}, which was detected in the ASCA Galactic Plane
Survey and is postulated based on its spectrum to be a candidate
supernova remnant \cite{bamba03:_diffus_hard_x_ray_sourc}, and is also
coincident with an H II region \cite{lockman89:HII_regions}. With its
angular size of $8.3'$ (FWHM), small distance (approximately 1.3 kpc),
and non-thermal spectrum, this object also may also contribute to a
component of the VHE emission. Additionally, the nearby source
{AX~J18406-0539} is possibly an XRB at a distance of 1.1 kpc
\cite{masetti06:INTEGRAL_1841}, though given positional errors, may
well be the same source as AX~J1841.0-0536 \cite{negueruela07:1841}.

\subsection{\object{\HESSg}}

\begin{figure*}[h]  
  \centering
  \resizebox{0.48\hsize}{!}{\includegraphics{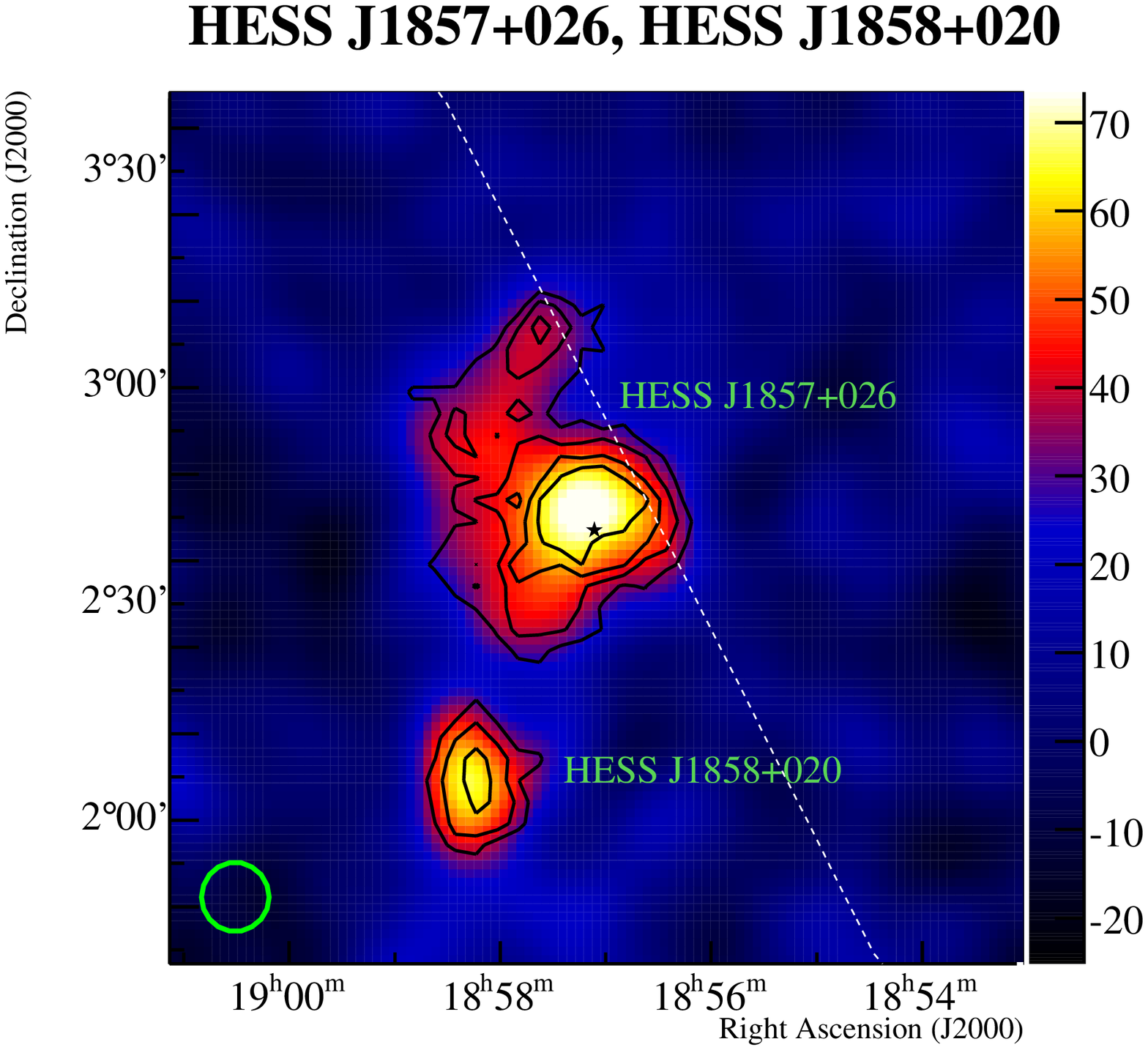}} 
  \resizebox{0.48\hsize}{!}{\includegraphics{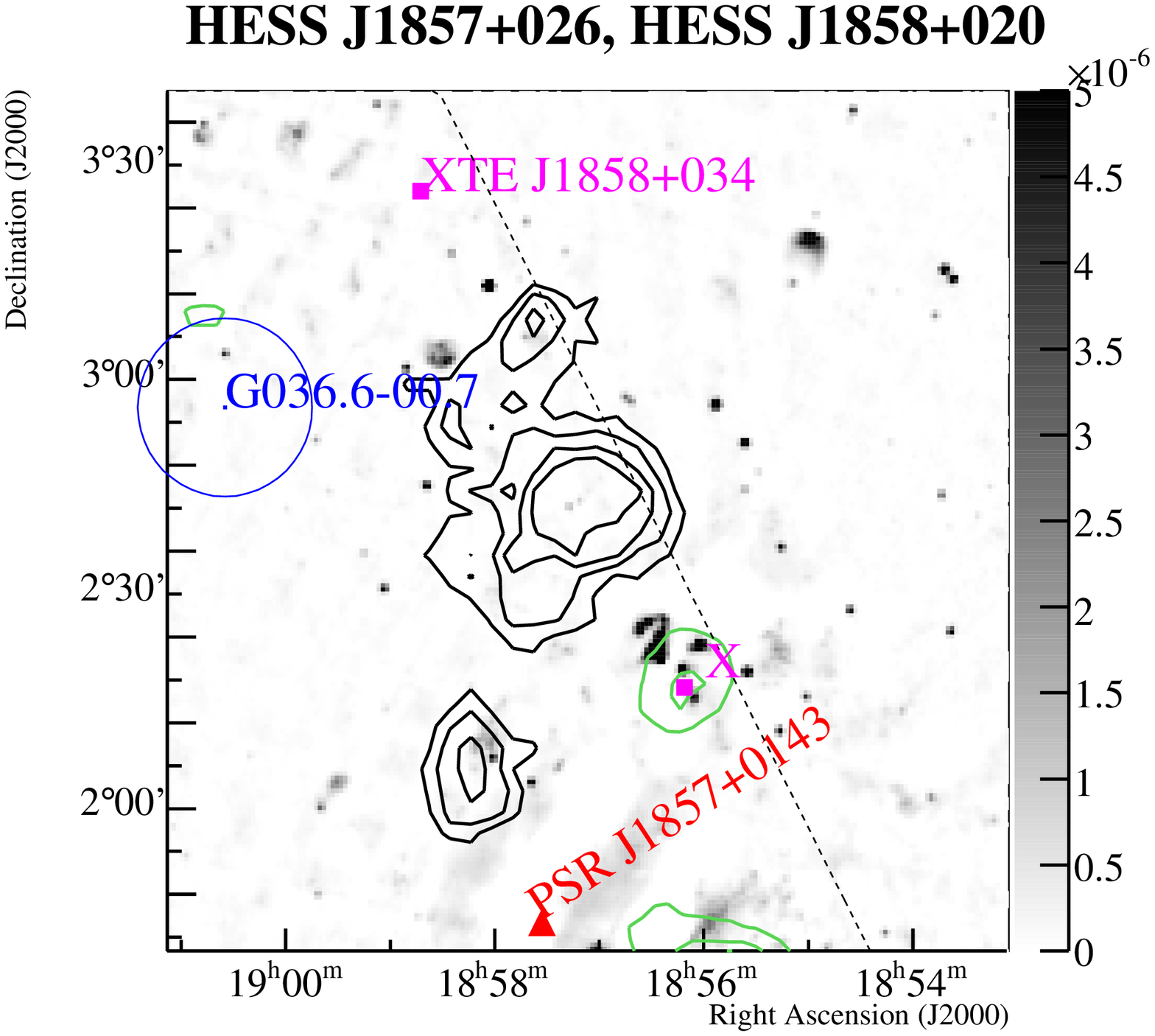}} 
  \caption{Left: A VHE gamma-ray image of \HESSg\ and \HESSh, plotted
    as in Figure \ref{fig:J1427}, with a  smoothing radius of
    $0.08^\circ$. \emph{Right:} The HESS significance
    (black) and adaptively smoothed ROSAT X-ray contours (green),
    overlaid on an NVSS radio image (grey-scale). Also shown are
    the positions of a known high spin-down flux pulsar (filled
    triangle), the SNR G036.6-00.7 (circle), the ROSAT point-source
    1RXS~J185609.2+021744 (labeled X), and the HMXB XTE~J1858+034
    (square).  }
  \label{fig:J1857}
\end{figure*}

\HESSg\ (Figure \ref{fig:J1857}) is an approximately
radially-symmetric extended VHE gamma-ray source located on the
Galactic Plane.  The source is detected by HESS at a $9\sigma$
significance level at energies above 300 GeV and has a differential
spectral index of \StatSysErr{2.4}{0.1}{0.2}. The slight extension of
the source seen toward the north is significant ($\sim 5\sigma$) and
may indicate a more extended morphology or the presence of a weaker
nearby source, though more observation time will be needed to make a
conclusive statement.

This source lies approximately 0.7$^\circ$ from \HESSh\ (see \S
\ref{sec:J1858}), which is most probably a separate source since no
significant emission connects the two. An association with the
supernova remnant G036.6-00.7, which lies over a degree from the
centroid position, is unlikely.  Though an ASCA observation exists
which is roughly centered on the source position, no excess was seen,
implying a 95\% absorbed flux upper-limit of
$1.2\cdot~10^{-12}\:\mathrm{erg\:cm^{-2}\:s^{-1}}$ between
2--10~keV. The X-ray source seen about a quarter of a degree from the
centroid position is the point-source {1RXS~J185609.2+021744}
(labeled $X$ in the Figure, and coincident with the ASCA source
{AXJ~185608+0218}), which has a flux in the 0.1--2.4 keV range
of $(0.32\pm0.06)\times 10^{-12}\:\mathrm{erg\:cm^{-2}\:s^{-1}}$,
assuming a photon index of 2.0; its distance from the emission region
makes it an unlikely counterpart candidate, however.

\subsection{\object{\HESSh}} \label{sec:J1858}


The weak gamma-ray source \HESSh\ (shown also in Figure
\ref{fig:J1857}) lies close to \HESSg; however, there is no
significant emission connecting them, suggesting that they are
distinct objects. It is detected at a significance level of $7\sigma$
with a differential spectral index of
\StatSysErr{2.2}{0.1}{0.2}. Though nearly point-like, its morphology
shows a slight extension of $\sim{\hskip -0.3em} 5'$ along its major
axis. {PSR~J1857+0143} ($\dot
E/D^2=1.7\cdot10^{34}\mathrm{erg\:s^{-1}\:kpc^{-2}}$) is powerful
enough to explain the source, but is significantly offset. 


\section{Summary}

The eight VHE gamma-ray sources discussed here are all extended
objects with angular sizes ranging from approximately 3 to 18 arc
minutes, lying close to the Galactic plane (suggesting they are
located within the Galaxy). In each case, the spectrum of the sources
in the TeV energy range can be characterized as a power-law with a
differential spectral index in the range 2.1 to 2.5.  The general
characteristics of these sources---spectra, size, and position---are
similar to previously identified galactic VHE sources (e.g. PWNe),
however since these sources have so far no clear counterpart in
\UPDATED{lower-energy} wavebands, further multi-wavelength study is required to
understand the emission mechanisms powering them, and therefore
follow-up observations with higher-sensitivity X-ray and GeV gamma-ray
telescopes will be beneficial.  Since most VHE sources are predicted
to emit X-ray and radio emission, a non-detection of \UPDATED{longer}-wavelength
emission with current-generation experiments for some of these objects
may be an indication that a new VHE source class exists \cite[as
suggested by][]{HESS:scanpaper1}, and may provide new insight into
high-energy processes within our Galaxy.

\begin{acknowledgements}

  The support of the Namibian authorities and of the University of
  Namibia in facilitating the construction and operation of
  HESS is gratefully acknowledged, as is the support by the German
  Ministry for Education and Research (BMBF), the Max Planck Society,
  the French Ministry for Research, the CNRS-IN2P3 and the
  Astroparticle Interdisciplinary Programme of the CNRS, the
  U.K. Science and Technology Facilities Council (STFC), the IPNP of
  the Charles University, the Polish Ministry of Science and Higher
  Education, the South African Department of Science and Technology
  and National Research Foundation, and by the University of
  Namibia. We appreciate the excellent work of the technical support
  staff in Berlin, Durham, Hamburg, Heidelberg, Palaiseau, Paris,
  Saclay, and in Namibia in the construction and operation of the
  equipment.

  This research has made use of the SIMBAD database, operated at CDS,
  Strasbourg, France and the ROSAT Data Archive of the
  Max-Planck-Institut f\"ur extraterrestrische Physik (MPE) at
  Garching, Germany.

\end{acknowledgements}

\bibliographystyle{aa} \bibliography{8516}

\end{document}